\documentclass[sigconf]{acmart}
\AtBeginDocument{%
  }

\copyrightyear{2026}
\acmYear{2026}
\setcopyright{cc}
\setcctype{by}
\acmConference[MobiCom '26]{The 32nd Annual International Conference on Mobile Computing and Networking}{October 26--30, 2026}{Austin, TX, USA}
\acmBooktitle{The 32nd Annual International Conference on Mobile Computing and Networking (MobiCom '26), October 26--30, 2026, Austin, TX, USA}
\acmDOI{10.1145/3795866.3844474}
\acmISBN{979-8-4007-2505-0/2026/10}

\usepackage{graphicx}
\usepackage{caption}
\usepackage{array}
\newcolumntype{L}[1]{>{\raggedright\arraybackslash}p{#1}}
\usepackage{caption}
\usepackage{algorithm}
\usepackage{algpseudocode}
\usepackage{amsmath}
\usepackage{xargs}
\usepackage[colorinlistoftodos,prependcaption,textsize=normalsize]{todonotes}
\usepackage{cleveref}
\usepackage[normalem]{ulem}
\usepackage{multirow}
\usepackage{makecell}
\usepackage{subcaption}
\usepackage{hyperref}
\usepackage{graphicx}
\usepackage{tabularx}
\usepackage{cleveref}
\usepackage{pifont}
\usepackage{xspace}
\usepackage{soul}
\usepackage{placeins}
\usepackage{subcaption}
\usepackage{lipsum}
\usepackage{dblfloatfix}
\usepackage{listings}
\usepackage{xcolor}
\usepackage{cuted}
\usepackage{afterpage}
\usepackage{placeins}

\lstdefinestyle{myjsonblock}{
    basicstyle=\ttfamily\scriptsize,
    breaklines=true,
    breakatwhitespace=true,
    columns=fullflexible,
    keepspaces=true,
    frame=single,
    showstringspaces=false,
    aboveskip=3pt,
    belowskip=3pt,
    moredelim=**[is][\color{red}]{@red@}{@endred@}
}

\newcommand{\cmark}{$\checkmark$} 
\newcommand{\cmarkbold}{\boldmath\checkmark}
\newcommand{\xmark}{$\times$}

\newcommand{\lm}{\textit{CellSecInspector}\xspace}
\newcommand{\core}{\textit{CellSecInspector-Core}\xspace}
\begin{document}

\title{\lm: Safeguarding Cellular Networks via Automated Security Analysis on Specifications}

\author{
Ke Xie\textsuperscript{1}, 
Xingyi Zhao\textsuperscript{1},
Min-Yue Chen\textsuperscript{2},
Yu-An Chen\textsuperscript{2},
Yiwen Hu\textsuperscript{3}, 
Munshi Saifuzzaman\textsuperscript{1}, \\
Wen Li\textsuperscript{1},
Shuhan Yuan\textsuperscript{1},
Guan-Hua Tu\textsuperscript{2},
Tian Xie\textsuperscript{1}
}

\affiliation{
\textsuperscript{1} Utah State University \quad
\textsuperscript{2} Michigan State University \quad
\textsuperscript{3} University of Maryland, Baltimore County \quad \\
\country{\{ke.xie, xingyi.zhao, munshi.saifuzzaman, awen.li, Shuhan.Yuan, tian.xie\}@usu.edu \\ \quad \{chenmi33, cheny132, ghtu\}@msu.edu \quad huyiwen@umbc.edu \quad}
}



\renewcommand{\shortauthors}{}

\begin{abstract}
\noindent The complexity, interdependence, and rapid evolution of 3GPP specifications present fundamental challenges for ensuring the security of modern cellular networks. Manual reviews and existing automated approaches, which often depend on rule-based parsing or small sets of manually crafted security requirements, fail to capture deep semantic dependencies, cross-sentence/clause relationships, and evolving specification behaviors. In this work, we present \lm, an automated framework for security analysis of 3GPP specifications. \lm extracts structured state–condition–action (SCA) representations, models mobile network procedures with comprehensive function chains, systematically validates them against 9 foundational security properties under 4 adversarial scenarios, and automatically generates test cases. This end-to-end approach enables the automated discovery of vulnerabilities without relying on manually predefined security requirements or rules. Applying \lm to the well-studied 5G and 4G NAS and RRC specifications and selected sections of TS 23.501 and TS 24.229, it discovers 43 vulnerabilities, 7 of which are previously unreported. Our findings show that \lm is a scalable, adaptive, and effective solution to assess 3GPP specifications for safeguarding operational and next-generation cellular networks. 

\end{abstract}




\maketitle

\section{Introduction}

\noindent Cellular networks are critical for modern global connectivity, with 5G and 4G forming the backbone of modern mobile communication systems. Technical specifications and architectures are standardized by the 3rd Generation Partnership Project (3GPP), an association of seven Organizational Partners, including ATIS in the U.S., ETSI in Europe, and CCSA in China. Thousands of specifications are published by 3GPP to ensure global interoperability across mobile networks and devices.
Consequently, mobile equipment vendors and operators are required to ensure compliance with 3GPP specifications. However, this standardization introduces systemic risks: 
\textit{Design flaws in 3GPP specifications can propagate into all compliant networks and devices, creating widespread threats with global negative impact}. 
Moreover, the increasing complexity, frequent revisions, and evolving new generations of mobile networks continue to expand attack surfaces. 

To mitigate those risks, researchers spend increasing effort to scrutinize cellular network specifications and services documented in the 3GPP specifications to identify design issues. However, the traditional manual analysis by experts is slow, error-prone, and infeasible at scale. 
State-of-the-art automated approaches primarily rely on rule-based approaches (e.g., dependency parsing \cite{chen2021bookworm,al2024hermes}), and keyword-driven scanning (e.g., hazard indicators~\cite{chen2021bookworm}), failing to capture the deeper and implicit knowledge in 3GPP specifications.  
As a result, vulnerabilities in critical mobile network procedures continue to be exploited, enabling attacks such as downgrade attacks~\cite{shaik2015practical, chen2021bookworm, 6550445, kambourakis2011attacks,
      lee2009detection, leong2014unveiling, hussain20195greasoner,
      hussain2018lteinspector, kim2019touching, 8894379}. 
These methods suffer from three fundamental limitations: (1) they lack deep insights into the logic, processes, and context of specification texts, particularly when dealing with cross-sentence/clause dependencies; (2) their security requirements are manually developed for specific procedures, which severely limits scalability across the thousands of 3GPP specifications; (3) their analysis pipelines require costly manual updates to remain synchronized with evolving 3GPP releases.  
These limitations highlight the urgent need for an intelligent, automated system capable of understanding 3GPP specifications and systematically assessing them.  

In this study, we develop \lm (\ul{Cell}ular Network \ul{Sec}urity \ul{Inspector}), an automated framework capable of understanding 3GPP specifications and systematically assessing them via foundational security properties. \lm models mobile networks in structured SCA (state-condition–action) representations from 3GPP specifications, and examines them with foundational security properties. By integrating specification analysis with automated test case generation, \lm delivers a scalable, adaptive, and future-proof solution for 3GPP specification security assessment. In this study, \lm analyzes the critical NAS and RRC layers for current operational 5G and 4G mobile networks as well as the selected sections of TS 23.501 and TS 24.229. 
\lm not only discovers 36 previously reported vulnerabilities but also reports 7 new vulnerabilities.
We summarize our contributions as follows. 

\noindent $\diamond$\textbf{New Framework:} We developed \lm, a novel approach for automated security assessment of 3GPP specifications, advancing toward fully automated and scalable specification analysis for safeguarding not only operational 5G/4G cellular networks but also Next-G mobile networks.  

\noindent $\diamond$\textbf{New Findings:} Applying \lm to 5G and 4G NAS and RRC specifications, as well as the selected sections of TS 23.501 and TS 24.229, we discovered 43 vulnerabilities, 7 of which were previously unreported. They are shown to have serious security consequences, demonstrating the urgent need for automated 3GPP specification analysis and the effectiveness of \lm in identifying known and previously undetected vulnerabilities. 

\section{Background}

\noindent \textbf{Non-Access Stratum (NAS) and Radio Resource Control (RRC):} The NAS is a key component of the control plane in modern 5G and 4G cellular networks, managing mobility, session control, and security between user equipment (UE) and the core network. It ensures seamless session establishment, authentication, and mobility across different networks. The RRC operates within the Access Stratum (AS) of both 5G and 4G networks and manages signaling between the UE and the radio access network (RAN). It handles connection setup, maintenance, release, handovers, paging, measurement reporting, and security configuration.

\noindent \textbf{Large Language Models (LLMs):} LLMs have emerged as powerful systems trained on massive text corpora to perform a wide range of natural language processing tasks, including comprehension, reasoning, and generation. Modern LLMs, including ChatGPT~\cite{brown2020language} and DeepSeek~\cite{bi2024deepseek}, have demonstrated impressive emergent abilities across diverse tasks, showing that LLMs can go beyond text processing to support domain-specific reasoning.
This capability makes LLMs a promising foundation for analyzing complex 3GPP specifications, where precision, contextual understanding, and reasoning across specifications are critical.

\section{Related Work}
\label{sect:related}

\noindent Given the scale and complexity of 3GPP specifications, NLP techniques have increasingly been applied for cellular network security analysis. 
\textit{Atomic}~\cite{chen2021bookworm} develops adversarial test cases from the 4G NAS specification. It relies heavily on keyword seeds (e.g., conditional words, risky operations) to identify relevant sentences to generate the test cases. 
\textit{ConTester}~\cite{chen2023sherlock} automates LTE NAS conformance test generation by constructing an Event Dependency Graph (EDG). 
Its scope is limited to generating test cases for assessing commercial mobile devices rather than analyzing design issues within 3GPP specifications.  
\textit{CellularLint}~\cite{rahman2024cellularlint} leverages a domain-adapted BERT~\cite{devlin2019bert} model to identify inconsistencies in 5G and 4G NAS and security-related specifications. Its emphasis is on sentence-level semantic conflicts, similar to natural language inference (NLI) tasks~\cite{dagan2005pascal,bowman2015largeannotatedcorpuslearning}, rather than systematic security analysis. 
\textit{Hermes}~\cite{al2024hermes} parses the specification via handcrafted grammars using constituency parsing and dependency parse trees to develop FSMs, which are further manually translated into SMV language~\cite{clarke1996symbolic,al2024hermes,hussain20195greasoner,hussain2018lteinspector} for formal security analysis. 
It remains rule-intensive and lacks flexibility to adapt to evolving specifications. \textit{ARCANE}~\cite{tan2025automated} leverages large language models to assist specification understanding and combines it with passive model learning to automatically construct 
FSMs for model-based fuzzing. The generated models are then used to produce fuzzing inputs for testing 5G O-RAN implementations. However, ARCANE focuses on implementation bug discovery rather than systematic analysis of design issues in 3GPP specifications. In contrast, \lm provides a security analysis framework that aims to comprehensively comprehend 3GPP specifications and conduct systematic analysis of cellular network specifications. It can be easily adapted to conduct security analysis of large-scale and evolving 3GPP specifications, providing an efficient and effective approach to complement traditional manual efforts. Table~\ref{tab:spec-compare} in Appendix \ref{append:Related_Work} compares these works.

\section{Overview of \lm}

\begin{figure*}
    \centering
    \includegraphics[width=0.82\linewidth]{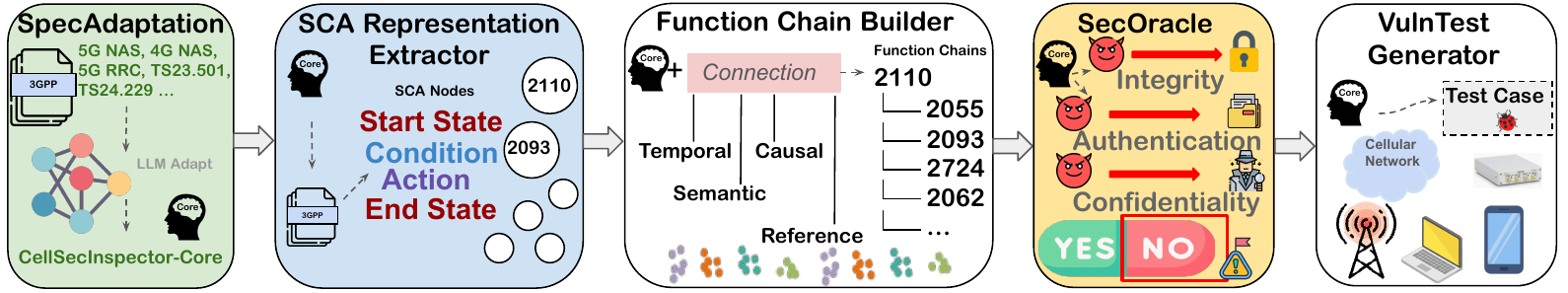}
    \vspace{-0.1in}
    \caption{\lm Overview}
    \label{fig:Overview}

\end{figure*}

\subsection{Approach Skeleton}
Figure~\ref{fig:Overview} shows the overview of \lm, which can automatically comprehend complicated 3GPP specifications, model the mobile network functions using structured state-condition-action (SCA) representations, and examine whether the 3GPP-designed procedures can violate foundational security properties. Five primary modules of \lm are developed: \textit{SpecAdaptation} (\ul{Spec}ification \ul{Adaptation}), \textit{SCA Representation Extractor}, \textit{Function Chain Builder}, \textit{SecOracle} (\ul{Sec}urity \ul{Oracle}), and \textit{VulnTestGenerator} (\ul{Vuln}erability \ul{Test} \ul{Generator}). 
\textit{SpecAdaptation} adapts the core model \textit{CellSec\-Inspector-Core} to the cellular network domain knowledge using the collected and curated 3GPP specifications. \textit{CellSecInspector\allowbreak-Core} is further used in later modules.
Two components are involved to comprehensively model the complicated mobile network functions. \textit{SCA Representation Extractor} parses and extracts the SCA nodes from 3GPP specifications and the internal domain knowledge learned via \textit{SpecAdaptation}. \textit{Function Chain Builder} connects scattered SCA nodes into function chains for comprehensively modeling the cellular network services specified in 3GPP specifications. Next, \textit{SecOracle} conducts the security checking using the foundational security properties. Lastly, \textit{VulnTestGenerator} automatically generates test cases for candidate violations.

\subsection{Challenges and Insights}
\label{subsect:challenges}

3GPP specifications are inherently complex and have a vast scope, complicating the development of a framework for comprehending the cellular networks governed by 3GPP specifications. For instance, NAS and RRC specifications span thousands of pages, with intricate interdependencies and frequent updates across releases. A single 5G NAS specification TS 24.501 contains over \texttt{981} pages~\cite{3GPP_TS_24501}. Reading and understanding 3GPP specifications is a time-consuming challenge even for experts. We design and develop \lm based on a pretrained LLM. 
To enable \lm to discover vulnerabilities from 3GPP specifications, we solve three main challenges.  

\noindent \textbf{C1: Lack of annotated resources for mobile network security research tasks.}
Different from common NLP (natural language processing) research tasks, such as text classification, that come with the thorough and well-annotated datasets, there are no well-known and easily accessible resources in the mobile network research community. Specifically, we need to model the complex mobile network procedures first to conduct the security analysis of mobile network functions and services. Unfortunately, there are no established and public annotation specifications or training resources in this domain-specific task. 

\noindent $\diamond$ \textbf{Insight1:} To address C1, we propose the \textit{SCA Representation Extractor}, which is designed to extract information for modeling the mobile network functions and services from 3GPP specifications. Two approaches are developed. First, instead of constructing the traditional FSM (finite state machine) models, \textit{SCA Representation Extractor} captures the self-contained transition of mobile network procedures in our proposed \textit{SCA (State-Condition-Action)} structure that can compose the complex and completed function event in a single presentation. Second, we adopt the efficient few-shot In-Context Learning (ICL) to enable our system to extract SCA nodes without large-scale annotated datasets. 

\noindent {\textbf{C2: Computational cost for linking SCA nodes at scale.} 
After deriving scattered SCA nodes, these nodes must be linked to construct complete mobile network procedures and capture the comprehensive network status (e.g., whether the security context is established). 

However, linking SCA nodes at scale is challenging because different procedures described in SCA nodes may or may not share meaningful relationships. 
Moreover, to avoid redundancy, 3GPP specifications use cross-references between clauses and specifications, further complicating the linkage process.  
With thousands or more SCA nodes that can create millions of candidate connections in a quadratic manner, it imposes substantial computational overhead for linking SCA nodes at scale. 

\noindent $\diamond$ \textbf{Insight2:} 
To address C2, we introduce the \textit{Function Chain Builder}, which integrates two approaches: Type I: Node-Informed Exhaustive Connection and Type II: Reference Guided Connection.
Type I compares all node pairs to identify three types of valid transitions: temporal connections that link nodes with identical end and start states, semantic connections that capture equivalent but textually different states, and causal connections where one node’s state, condition, or action enables another. While Type I provides comprehensive coverage, its $O(n^2)$ complexity is costly as the number of nodes grows. To reduce this burden, Type II leverages the rich cross-references embedded in 3GPP specifications (e.g., `\textit{as specified in subclause X.Y.Z}') to restrict the search space to relevant sections. Together, these strategies enable efficient and automated construction of function chains across 3GPP specifications.

\noindent \textbf{C3: Manually crafted security requirements are impractical for security assessment of large-scale mobile network functions.} 
State-of-the-art methods, such as \textit{Hermes}~\cite{al2024hermes}  and \textit{ConTester}~\cite{chen2023sherlock}, rely on a small set of manually crafted security  requir-\ ements for specific procedures. For example, ConTester’s security requirement S6 mandates that `\textit{Except the messages listed below, no NAS signaling messages shall be processed by the receiving EMM entity in the UE forwarded to the ESM entity, ...}', which is the sentence directly from TS 24.301 subclause 4.4.4.2. When new NAS procedures are introduced in subsequent 3GPP releases, if these sentences are updated, their security requirements must be manually revised, limiting the scalability for the security assessments of large-scale mobile network functions. Similarly, approaches that build FSM models also need to manually translate FSM models into SMV language~\cite{clarke1996symbolic,al2024hermes,hussain20195greasoner,hussain2018lteinspector} to enable formal security analysis, requiring significant expert efforts that are not practical given the complexity and breadth of modern mobile networks.  

\noindent $\diamond$ \textbf{Insight3:} We develop the \textit{SecOracle} that introduces 9 foundational security properties. 
\textit{Unlike procedure-specific requirements, these properties are derived from classic and foundational security principles applicable across the entire mobile network stack. }
Therefore, \textit{SecOracle} can operate on large-scale cellular network procedures with broader security checkpoints. This equips \lm with the foundational ability to detect previously unreported vulnerabilities.
\section{Detailed Design}
\label{sect:design}

We next present \lm in detail. Additional implementations and formalized algorithms are provided in Appendix~\ref{append:prompt_algorithm}.

\subsection{\textit{SpecAdaptation}}
We develop the core reasoning engine, \core, on top of open-source LLMs, in \textit{SpecAdaptation} to capture the cross-sentence and cross-clause relationships from the complex 3GPP specifications. To enable it to generalize and reason in 3GPP specifications, we expose \core to broad information to obtain cellular specification domain knowledge in the following two steps. 

\noindent \textbf{Cross-sentence/clause relation preserving data sanitation.} 
To ensure \core can be exposed to the broad cellular network domain knowledge, we begin by collecting and curating 5G and 4G technical specifications from the 3GPP portal~\cite{3GPP_Specs_2025} as the corpus for training. To ensure consistency and clarity, we sanitize the specifications by (a) removing redundant spaces, extraneous line breaks, and leading symbols (e.g., dots, bullets, hyphens), (b) trimming trailing punctuation, (c) discarding empty parentheses and special unicode characters, and (d) removing tables and figures, which are usually explained or demonstrated in texts. Note that \ul{references (e.g., `as specified
in subclause X.Y.Z') are preserved in the datasets}, ensuring \core can learn the cross-sentence and cross-clause relations via explicit references that are neglected by state-of-the-art works~\cite{rahman2024cellularlint}.

\noindent \textbf{Domain knowledge adaptation.} A straightforward approach to expose standards knowledge to \core is to continually train (i.e., pretrain) the model on 3GPP specifications. While feasible in principle, this approach has two key limitations. 
First, pretraining a model is computationally expensive and difficult to maintain for the fast-evolving standards. 3GPP standards evolve frequently across different releases and versions. A continuously trained model can become out-of-date quickly unless training is performed repeatedly. Second, training an LLM on a narrow domain corpus (i.e., specifications) under limited compute and data diversity can degrade the model's general reasoning and instruction-following abilities~\cite{gururangan2020don, kemker2018measuring, ouyang2022training}. For example, a continually pretrained model may exhibit catastrophic forgetting and spec-confusion behaviors, such as conflating procedures across different releases, hallucinating missing steps, or recurring boilerplate in specifications. These failures are harmful for standards analysis, where correctness hinges on precise conditions, message ordering, and release-specific normative requirements.

In our design, we leverage Retrieval-Augmented Generation (RAG)~\cite{lewis2020retrieval} to adapt \core to domain knowledge without modifying the base model. Specifically, \textit{CellSec\-Inspector-Core} retrieves relevant specification excerpts and conditions its generation on them, using the retrieved text as grounded evidence to complement its parametric knowledge. 
The retrieval corpus is organized as versioned specification chunks, where each chunk preserves the source specification number, release/version, clause identifier, and cross-reference information.
Concretely, given a query (e.g., message type, timer, or a security requirement), it searches a versioned corpus of 3GPP documents, which is prepared in the aforementioned step, and attaches the top-matched paragraphs (including section and release/version identifiers) to the model prompt. This retrieval step provides two technical benefits: (i) it ensures that \core uses the explicitly selected release/version text, and (ii) it reduces hallucination by requiring outputs to be justified by retrieved normative statements rather than relying solely on the model’s memory. Finally, \core obtains the 3GPP specification domain knowledge, and it will be further used in all modules. The detailed workflow is provided in Algorithm~\ref{alg:rag_prompt} in Appendix~\ref{append:prompt_algorithm}.

\subsection{\textit{SCA Representation Extractor}}
\label{subsect:sca_representation}

To address C1 and comprehensively model the complicated cellular network functions, we develop the \textit{SCA (State-Condition-Action) Representation Extractor} that extracts the structured function representations in SCA nodes from 3GPP specifications. Each SCA node contains four fields: the start state, representing the initial state before the transition; the condition, denoting the triggering clause or prerequisite specified in the text; the action, describing the operation mandated by the specification upon satisfaction of the condition; and the end state, indicating the resulting state after the action is executed. 

\noindent\textbf{SCA node vs. FSM state.} Different from the traditional finite state machine (FSM) that separates the states and transitions and stores the transition logic on edges between states, the SCA representation structure (i.e., node) aims to directly capture the self-contained transition, binding the logic of condition and action. This structure introduces two main advancements to maximize the utilization of the text processing capability of \core. 
First, it presents explicit transition semantics, including where the transition starts, when it is valid (condition), what happens (action), and where it ends. Each SCA node is one self-contained transition. 
The explicit viewpoint of a transition saves the step of merging all involved edges in the FSM when transitioning from one state to another.  
That is, an FSM state may have several edges with its connected states, requiring extra effort to identify how to transition to the next state. 
Second, it allows complex conditions and actions, combining internal variables, timers, and status, to be described in a single node, which is valuable for modeling cellular network functions where the transitions may depend on complicated conditions and actions. \textit{Note that SCA nodes can be easily transformed into the FSM states and edges. However, it is challenging to transform them in reverse.}

\noindent\textbf{Few-shot training with minimal dataset development overhead.}  
To extract SCA nodes from 3GPP specifications, instead of developing a large dataset of many SCA nodes training cases with heavy manual engineering overhead, we leverage efficient in-context learning (ICL)~\cite{brown2020language} to educate \lm without time-consuming fine-tuning to update model parameters.
ICL works by embedding carefully selected demonstration examples and task-specific instructions directly into the model’s input context, allowing the model to infer the desired output format. This method enables \lm to generalize to the SCA extraction task with a limited number of examples. 
Specifically, we instruct \lm to extract structured specification events from 5G and 4G specifications by mapping each relevant sentence into an SCA node. 
Along with the instruction, a few manually developed SCA nodes are given for \lm to infer. From each specification, we develop two examples to demonstrate common specification patterns. 
Sentences that accurately describe function events are expected to populate all four fields. 
If a sentence contains cross-references (e.g., `as specified in subclause X.Y.Z'), the referenced content is retained verbatim in the most relevant field. For ambiguous or underspecified sentences, we employ best-effort inference. If no definition is found, the corresponding field is marked as `Not explicitly defined' to ensure completeness of the extracted structure. Figure~\ref{fig:sca_extractor_prompt} in Appendix~\ref{append:prompt_algorithm} shows an example ICL instruction.

\begin{figure}[t]
  \centering
  \includegraphics[width=0.85\linewidth]{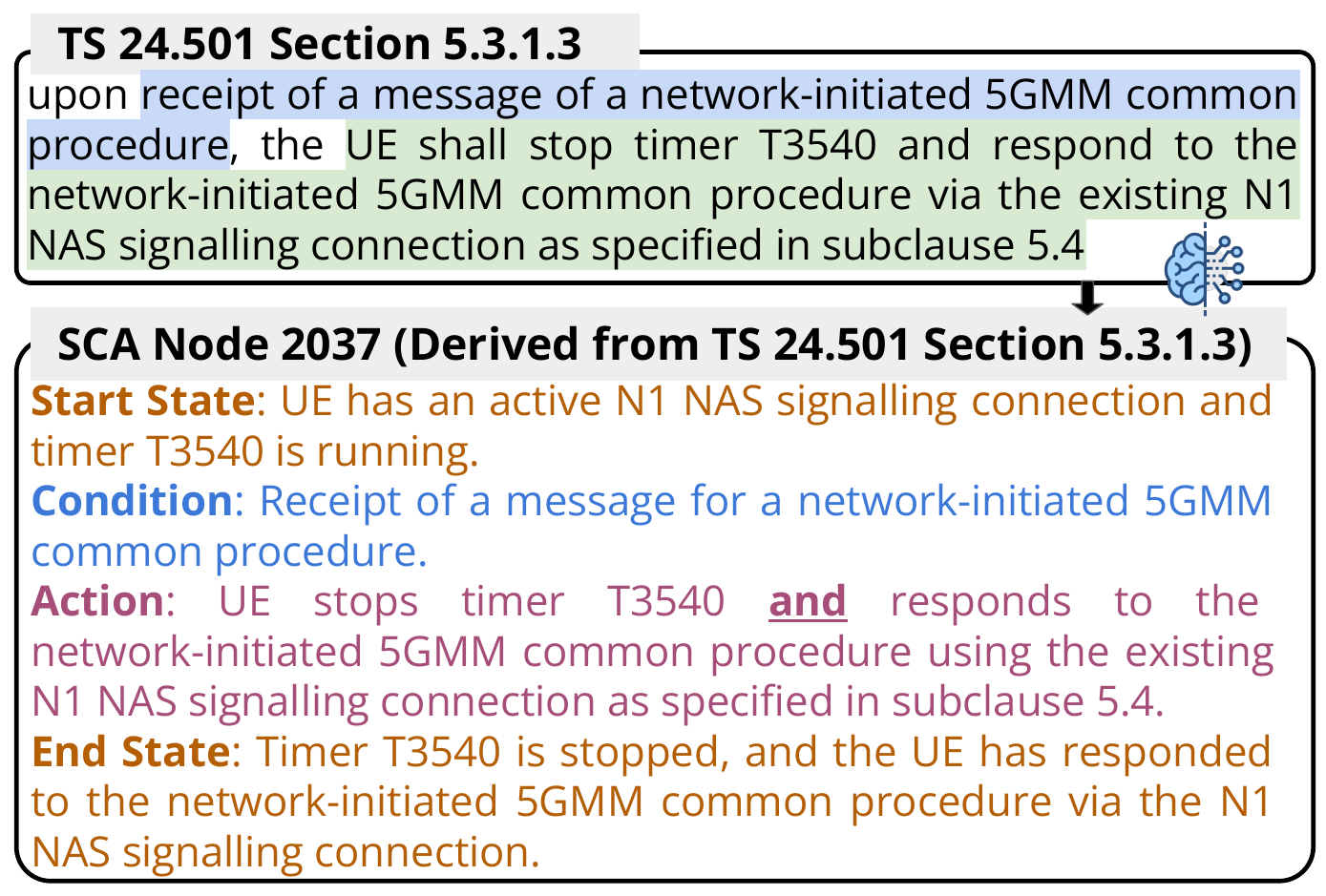}
  \vspace{-0.1in}
  \caption{SCA representation for Node 2037}
  \label{fig:SCAnode} 
  \vspace{-0.2in}
\end{figure}

\noindent\textbf{SCA node example.} 
Figure~\ref{fig:SCAnode} presents an extracted SCA node from a short description in 5G NAS specification TS 24.501 by the following steps.       
\lm first parses the sentence structure using domain knowledge learned in \textit{SpecAdaptation}. There was no explicit start state in this given example. \lm identifies the triggering condition by detecting the phrase upon receipt of a message, recognizing that the transition is activated when the UE receives a message for a network-initiated 5GMM common procedure. Next, \lm extracts the actions from the phrases `\textit{shall stop timer T3540}' and `\textit{respond to the procedure},' which are organized as the structured expression that covers these two operations: stop T3540 \textbf{and} respond via the existing N1 NAS signalling connection as specified in subclause 5.4. From the derived condition and action, including `\textit{via the existing N1 NAS signalling connection}' and `\textit{timer T3540}', it is inferred that the UE already maintains an active N1 NAS signalling connection and T3540 is running before the event occurs, which are captured as the start state. Last, \lm derives the end state according to the start state, condition, and actions: after stopping T3540 and responding, the UE remains connected via the N1 NAS signalling connection, but T3540 is no longer running. In a traditional FSM, it is impractical to infer the start and end states, and two operations in the action may bring unnecessary difficulties in determining which edges shall be included within complicated FSM transitions when there are thousands of connected states.

\subsection{\textit{Function Chain Builder}}
After extracting scattered SCA nodes, each representing a self-contained state transition, we assemble them into coherent function chains, which complete the transitions for representing the complex mobile network functions specified in 3GPP specifications. Notably, this task is non-trivial due to the challenge outlined in \textbf{C2} (Section~\ref{subsect:challenges}), where thousands or more SCA nodes can be developed from thousands of pages from a single 3GPP specification. To construct complete function chains from these scattered SCA nodes, two types of connection strategies are designed:

\subsubsection{Type I: Node-Informed Exhaustive Connection}
~This category uses exhaustive pairwise comparisons between SCA nodes to discover all valid transitions. Three connection methodologies are defined:

\noindent \textbf{(a) Temporal Connection.} 
A temporal connection captures the most basic dependency between SCA nodes. It identifies whether the end state of an SCA node is identical to the start state of another. Formally, let the set of extracted SCA nodes be $N = \{n_1, n_2, \ldots, n_m \}$, where each node $n_i$ is defined as   
\[
n_i = (\text{start}(n_i), \text{condition}(n_i), \text{action}(n_i), \text{end}(n_i)).
\]
\noindent Each node $n_i \in N$ is associated with a start state and an end state that $\text{start}(n_i), \; \text{end}(n_i) \in Q$, where $Q$ denotes the set of all specification states extracted from the 5G and 4G NAS and RRC specifications. 
The Temporal-Connection relation \( T \) is defined as 
\(T = \{ (n_i, n_j) \in N \times N \mid \text{end}(n_i) = \text{start}(n_j) \} \). 
SCA node \(n_j\) is a valid temporal successor of \(n_i\) if the start state of \(n_j\) is identical to the end state of \(n_i\). 
The formalized algorithm is shown in Algorithm~\ref{alg:temporal_connection} in Appendix~\ref{append:prompt_algorithm}.
As shown in Figure~\ref{fig:eventchain}, the SCA Node 2110 extracted from the 5G NAS specification presents a transition from 5GMM-IDLE mode with suspend indication to 5GMM-CONNECTED mode upon receiving an indication from lower layers that the RRC connection has resumed. Another Node 2055 represents a transition starting exactly in 5GMM-CONNECTED mode, describing the transition from 5GMM-CONNECTED to 5GMM-CONNECTED with RRC inactive. Because the end state of Node 2110 is identical to the start state of Node 2055, they meet the requirement for a valid temporal connection. 

\noindent \textbf{(b) \textbf{Semantic Connection.}} Temporal Connection requires that two SCA nodes must have exactly the same end state and start state. However, 3GPP specifications are written by different experts. There may be slight differences in textual descriptions for the identical concepts or functions in 3GPP specifications, which we refer to as \textit{textually different but semantically equivalent descriptions}.
To connect SCA nodes with such a relation, we introduce Semantic Connection, which identifies if the end state \(\text{end}(n_i)\) of Node $n_i$ and the start state \(\text{start}(n_j)\) of Node $n_j$ are semantically equivalent or strongly related. If so, Node $n_j$ is allowed to be the successor of Node $n_i$, even when the textual descriptions of \(\text{end}(n_i)\) and \(\text{start}(n_j)\) differ. Formally, given two SCA nodes, we check whether their semantics are $sem(\text{end}(n_i),\ \text{start}(n_j)) = \text{Yes}$ 
to associate the SCA nodes with \textit{textual descriptions different but semantically equivalent} for their end states and start states. In our approach, we leverage \core, which has been trained with extensive domain knowledge, to compare their semantic meaning. It takes 3 steps: (i) the SCA nodes' start and end states are tokenized; (ii) the transformer layers are used to build contextual embeddings based on tokens; (iii) the embeddings are further compared via cosine similarity and enhanced verification via ICL to distinguish deeper logic relations, such as entailment, contradictions, and overlaps. 
This relation \( S \)
is defined as
\[
S = \left\{(n_i, n_j) \in N \times N \;\middle|\; sem(\text{end}(n_i),\ \text{start}(n_j)) = \text{Yes} \right\} 
\]
The formalized algorithm is illustrated in Algorithm~\ref{alg:semantic_connection} in Appendix~\ref{append:prompt_algorithm}.
As shown in Figure~\ref{fig:eventchain}, Node 5263 ends in a state where the UE is in 5GMM-IDLE mode and has initiated the registration procedure due to a change in its radio capability.  
Node 2107 begins in a state where the UE is attempting to send a REGISTRATION REQUEST message. 
It proceeds to enter 5GMM-IDLE mode and continues the registration procedure. Although the end and start states are described differently, they both refer to the same specification state, the UE initiating a registration procedure triggered by updated radio capability. This represents a valid semantic connection. 

\begin{figure}
    \centering
    \includegraphics[width=0.86\linewidth]{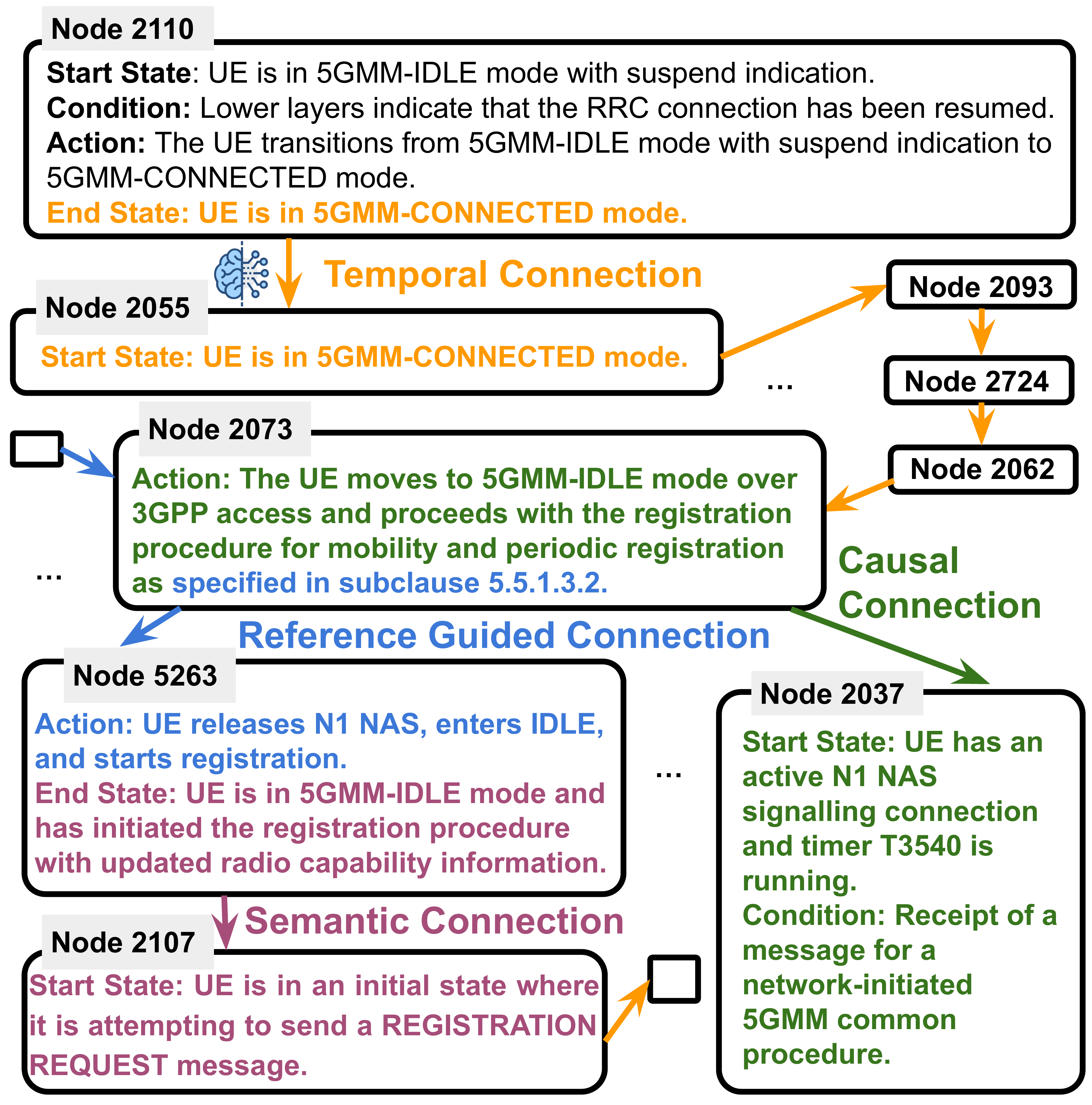}
    \vspace{-0.1in}
    \caption{5G mobility management function chain}
    \label{fig:eventchain}
    \vspace{-0.1in}
\end{figure}

\noindent \textbf{(c) \textbf{Causal Connection.}}
Beyond the direct matching of end and start states, function chains can be causally constructed through causal relationships, where the states, conditions, or actions of one SCA node directly enable, influence, or trigger the states, conditions, or actions of another.  
That is, the states, conditions, or actions in $n_i$ serve as a plausible prerequisite or cause for the states, conditions, or actions in $n_j$, suggesting a likely sequential or dependent relationship. 
Identifying causal connections between nodes is essential to thoroughly construct complete and coherent function chains. 
We define this Causal-Connection relation $C$ such that for any pair of SCA nodes $(n_i, n_j)$, a connection exists if $ C = \{ (n_i, n_j) \in N \times N \, |\ \begin{array}{l}
\text{causal}(n_i, n_j) = \text{Yes}
\end{array}
\}$. 
The formalized algorithm is shown in Algorithm~\ref{alg:causal_connection} in Appendix~\ref{append:prompt_algorithm}.
\core is employed to infer such causal relations. 
It analyzes SCA nodes $n_i$ and $n_j$ and determines whether $n_i$ precedes $n_j$. For example, if the end state of $n_i$ transitions the UE into a mode that is required as the start state of $n_j$, or if the condition in $n_i$ (e.g., radio capability change) leads to a logically subsequent condition in $n_j$ (e.g., registration request trigger), or if the action in $n_i$ (e.g., releasing NAS signaling connection) sets up the environment necessary for the action in $n_j$ (e.g., initiating a new signaling procedure), a causal relationship is inferred.

As shown in Figure~\ref{fig:eventchain}, Node 2073 models the UE transitioning from 5GMM-CONNECTED to 5GMM-IDLE mode and initiating a registration procedure. As part of this process, the UE establishes an N1 NAS signalling connection, which is necessary for subsequent communication with the network. 
Node 2037 begins in a state where such a signalling connection is already active and the UE is ready to handle network-initiated 5GMM common procedures. 
Its condition, receiving a message from the network, relies on the existence of an active signaling context. 
This requirement is fulfilled by the action in Node 2073, which creates the signaling connection through registration. With this context, Node 2037 can further proceed to receive or respond to the network-initiated procedure. 
Therefore, the action in Node 2073 is a prerequisite for the state and condition of Node 2037. We identify it as a valid causal connection $(n_{2073}, n_{2037}) \in C$.  

\subsubsection{Type II: Reference Guided Connection} 
~While the connection strategies in Type I can, in principle, be applied to all SCA nodes, this approach is computationally expensive as discussed in \textbf{C3} (see Section~\ref{subsect:challenges}).  
A single 3GPP specification can yield thousands of nodes, and across the more than 2,000 specifications that define operational 5G and 4G mobile networks~\cite{3gppspecs}, exhaustive pairwise validation is infeasible. 
To reduce the computational cost, we constrain the search space so that each SCA node is only checked with other nodes within the same subclause. However, this restriction risks constructing incomplete function chains. 
We introduce the Reference Guided Connection to address it. 
To ensure consistency and maintainability, 3GPP specifications are written in a highly modular style, relying heavily on cross-references to avoid redundancy and to coordinate contributions across multiple working groups. 
These references are the key to Reference Guided Connection, which works as follows: 
\noindent \textbf{1. Reference Detection}: Identify whether an SCA node \( n_i \) contains references to specific clauses (e.g., `\textit{as specified in subclause X.Y.Z}'). 
\textbf{2. Space Expansion}: Extract the referenced clause and retrieve all candidate nodes \( n_j \) within that clause. 
\textbf{3. Connection Validation}: Apply Type I connections to determine if a valid connection exists between \( n_i \) and \( n_j \).
The formalized algorithm is shown in Algorithm~\ref{alg:reference_connection} in Appendix~\ref{append:prompt_algorithm}.

As illustrated in Figure \ref{fig:eventchain}, we apply the Reference Guided Connection strategy to link Node 2073 and Node 5263. First, Node 2073 explicitly cites subclause 5.5.1.3.2 in its action field, which specifies the registration procedure triggered by radio capability updates. Based on this reference, all SCA nodes derived from subclause 5.5 are considered candidate targets, including Node 5263.
We then apply Type I mechanisms to validate whether a true dependency exists. In this instance, a Causal Connection from Node 2073 to Node 5263 is detected. Specifically, Node 2073 initiates a registration procedure. This behavior corresponds to the more detailed process described in Node 5263, where the registration is triggered by a radio capability change. The action in Node 2073 logically precedes and leads to the action in Node 5263, whose execution realizes the registration behavior initiated in Node 2073. Although their start and end states differ, the procedural flow from Node 2073 to Node 5263 is a causal dependency. Thus, it is a valid Reference Guided Connection.

\subsubsection{Complete Function Chain}~Applying all proposed connection strategies to analyze the relations among SCA nodes, the \textit{Function Chain Builder} can efficiently and comprehensively construct the function chain. 
These function chains serve as structured models of complex mobile network services. Figure~\ref{fig:eventchain} depicts the constructed function chains that model 5G mobility management.   

\subsection{SecOracle}
\label{subsect:secoracle}

\textit{SecOracle} (\ul{Sec}urity \ul{Oracle}) is an adaptive security validation module to assess the security of procedures and functions defined in 3GPP specifications. Current methods, such as exhaustive character-altered testing (i.e., random fuzzing~\cite{godefroid2020fuzzing}), manually developing mobile network procedure-specific requirements~\cite{chen2023sherlock,al2024hermes}, or manually translating FSM models into SMV language~\cite{clarke1996symbolic,al2024hermes,hussain20195greasoner,hussain2018lteinspector}, are inefficient and unscalable for assessing large-scale function chains (\textbf{C2} in Section~\ref{subsect:challenges}). 

To address this, we propose a two-step approach: 
First, we define a set of security properties, which are different from the security requirements in prior works that are specific to each procedure or FSM model~\cite{chen2023sherlock,al2024hermes}. \ul{These properties encapsulate the fundamental requirements for entire mobile networks, encompassing both mobile users and network infrastructure operators.} They thus can cover a wider range of scenarios and conditions. 
Derived from security standards (e.g., 3GPP Security Assurance Methodology (SECAM)~\cite{3gppTR33916v16}, ITU-T X.805~\cite{ituTX805}) and classic security models (e.g., Dolev-Yao~\cite{dolev2003security}, Bell-LaPadula~\cite{lapadula1996secure}, and Biba Models~\cite{Biba}), we build 9 core security properties in total (see Appendix~\ref{append:securityproperties}) to cover key security dimensions. The followings are a partial of our defined security properties: \texttt{authentication}, ensuring only legitimate users or entities can initiate procedures; \texttt{authorization}, enforcing that state transitions, such as NAS transitions, are exclusively triggered by legitimate, authenticated entities; \texttt{service integrity and confidentiality}, requiring all communications to be properly secured, with mobile entities verifying message integrity and confidentiality to prevent adversaries from launching attacks such as MitM. 

Second, motivated by the fact that cellular network devices, including UEs, base stations, and function modules in Core Network, do message exchanges all the time, 
4 fundamental attack methods, including dropping, modifying, injecting, and replaying, denoted as $A$ are applied to each node and node transition. \textit{SecOracle} leverages its reasoning capability to systematically analyze whether function chains still meet all security properties under these 4 attacks. Formally, given $\forall R_i \in $Security Properties, $\forall n_j \in \text{C}$ for each function chain, $\forall a_k \in \text{A}$,  the security property verification is expressed as, 
\(
\text{CheckViolation}(R_i, n_j, a_k).
\)
A candidate violation is flagged and recorded as 
\(
C, a_k, n_j \models \neg R_i
\) 
for further analysis. 
Figure~\ref{fig:secoracle} in Appendix~\ref{append:prompt_algorithm} shows how \core is instructed in this module.

\noindent \textbf{From candidate violations to confirmed vulnerabilities.}  
\textit{SecOracle} outputs candidate violations rather than confirmed vulnerabilities. Both candidate violations and confirmed vulnerabilities refer to \textit{design defects} in 3GPP specifications. A candidate violation indicates that a function chain may violate a security property under an adversarial action and therefore requires further validation. We classify a candidate as a confirmed vulnerability only if it satisfies three criteria: (i) the flagged violation is grounded in the corresponding 3GPP specifications; (ii) the issue reflects a specification-level design weakness rather than only an implementation-specific behavior; and (iii) the issue can lead to a realistic security consequence, such as unauthorized state transition, service disruption, privacy leakage, or inconsistent UE/network state.

\noindent \textbf{Violation example.} Consider the function chains shown in Figure~\ref{fig:eventchain}. For each node, \textit{SecOracle} applies 4 attack methods to assess possible candidate violations against security properties. In Node 2073, the UE sends a REGISTRATION REQUEST to transition from 5GMM-CONNECTED to 5GMM-IDLE mode. As a detected candidate violation shown in Figure~\ref{fig:output_secoracle}, under a replay attack, if this REGISTRATION REQUEST is intercepted and replayed, the network may process it again. According to 3GPP 5G NAS specification TS 24.501 subclause 4.4.4.3~\cite{3GPP_TS_24501}, the AMF must accept such a message even if it is unauthenticated and lacks integrity protection. 
The acceptance of an unauthenticated initial REGISTRATION REQUEST is a bootstrapping mechanism in a zero-trust environment. However, as reported in GSMA CVD\#0029~\cite{hussain20195greasoner}, the AMF may disconnect an existing legitimate UE connection after receiving an unauthenticated initial REGISTRATION REQUEST, before the requesting entity has been sufficiently authenticated or verified. In this case, the attacker can exploit the fact that an unauthenticated request can trigger security-sensitive connection-state changes. These inconsistencies can lead to billing errors, misrouted downlink traffic, or even session management failures. In severe cases, adversaries can repeatedly force the UE to re-register, thereby disrupting ongoing sessions and achieving a denial-of-service (DoS) attack using forged NAS messages. This is a specification design vulnerability; the violated property is \texttt{service integrity}: security-sensitive actions, such as releasing an existing UE connection or replacing an active UE context, should not be triggered solely by an unauthenticated message before sufficient verification.

\begin{figure}[t]
\centering
\begin{minipage}{0.98\columnwidth}
\begin{lstlisting}[style=myjsonblock]
| 2062 | 2073 | Replay | BEGIN_JSON {
  "vulnerability_detected": "Yes",
  "violated_property": [5],
  "explanation": "A Replay attack could exploit the transition from 5GMM-CONNECTED mode with RRC inactive indication to 5GMM-IDLE mode by capturing and retransmitting the REGISTRATION REQUEST message. This violates the Replay Attack Protection requirement.",
  "test_case": "An attacker captures the REGISTRATION REQUEST message during the transition and retransmits it, causing the AMF to process the same message twice, leading to potential registration failures or security breaches.",
  "vulnerability_source": "event2073"
} END_JSON |
\end{lstlisting}
\end{minipage}
\vspace{-0.1in}
\caption{A candidate violation flagged by \textit{SecOracle} for Nodes 2062 and 2073 transition}
\label{fig:output_secoracle}
\vspace{-0.1in}
\end{figure}

\subsection{VulnTestGenerator}

After detecting 
candidate violations by \textit{SecOracle}, \textit{VulnTestGenerator} generates corresponding testing procedures for validating whether these candidates can be confirmed as vulnerabilities.  
It aims to narrow the gap between specification-level findings and real-world verifications. As shown in Figure~\ref{fig:vulnTestGenerator} in Appendix~\ref{append:prompt_algorithm}, \textit{VulnTestGenerator} is instructed via ICL to develop structured test cases that specify network states, security context status, UE and core network configurations, operation sequences, and expected outputs, from flagged candidate violations. Those test cases can guide us how to validate and observe the violated security properties. 
Table \ref{tab:replay-2073} in Appendix~\ref{append:testcase} presents an example test case for validating the security property \texttt{service integrity} violation in Node 2073,  
which describes a specific behavior in TS~24.501 Subclause~5.3.1.4, where a UE in 5GMM-CONNECTED mode with RRC inactive indication sends a REGISTRATION REQUEST with the NG-RAN-RCU bit set for UE radio capability update and then moves to 5GMM-IDLE.

\noindent \textbf{Validating violations with test cases.} Executing the generated test cases on either a controlled testbed (e.g., open5GS~\cite{open5gs}, srsRAN-\ ~\cite{srsran}) or an operational mobile network allows us to validate the vulnerability and find the root cause. Notably, conducting real-world experiments requires substantial human effort, including building experimental platforms, instrumenting the network/UE to collect experimental data, and analyzing results. The current scope of \textit{VulnTestGenerator} focuses on generating test cases to guide these validation experiments. Executing test cases in an automated manner is a non-trivial activity. We leave it as future work.

\section{Evaluation}

\noindent This section reports the evaluation of \lm. Note that all datasets, scripts, and findings are available on the project website~\cite{cellsecinspector2025}. Implementations and 3GPP specifications used for evaluations are introduced in Appendix~\ref{append:implementation-datasets}. 

\subsection{Research Question}
To evaluate the performance of \lm, we answer the following research questions. 

\noindent $\diamond$ \textbf{RQ1:} Can \lm identify new vulnerabilities in 3GPP specifications?  \\
\noindent $\diamond$ \textbf{RQ2:} How does \lm compare with the state-of-the-art approach in vulnerability discovery effectiveness? \\
\noindent $\diamond$ \textbf{RQ3:} How effectively does SCA preserve procedural information? \\
\noindent $\diamond$ \textbf{RQ4:} What's the conversion rate from candidate violations to confirmed vulnerabilities?\\
\noindent $\diamond$ \textbf{RQ5:} What is \lm's runtime performance?

\subsection{RQ1: New Vulnerabilities}
\label{subsect:rq1_new_vulnerabilities}
\begin{table*}[t]
\centering
\caption{Summary of new vulnerabilities }
\vspace{-0.1in}
\label{tab:new_vulnerabilities}
\scriptsize

\scriptsize
\setlength{\tabcolsep}{2pt}
\renewcommand{\arraystretch}{1.05}

\begin{tabular}{
|>{\raggedright\arraybackslash}p{0.9cm}
|>{\raggedright\arraybackslash}p{2.2cm}
|>{\raggedright\arraybackslash}p{4.1cm}
|>{\raggedright\arraybackslash}p{2.0cm}
|>{\raggedright\arraybackslash}p{2.55cm}
|>{\raggedright\arraybackslash}p{4.5cm}|
}

\hline
\multicolumn{1}{|c|}{\multirow{2}{*}{\textbf{Spec\#}}}
& \multicolumn{1}{c|}{\multirow{2}{*}{\textbf{Vulnerability}}} 
& \multicolumn{1}{c|}{\multirow{2}{*}{\textbf{Description}}} 
& \multicolumn{2}{c|}{\textbf{PoC Validation}} 
& \multicolumn{1}{c|}{\multirow{2}{*}{\textbf{Root cause}}} \\
\cline{4-5}
& & 
& \textbf{Validation Setup}
& \textbf{Experimental Result}
& \\
\hline

\multirow[t]{2}{=}{\raggedright
TS 24.501\newline
v16.8.0~-- \newline
v20.0.0$^\dagger$}
& V1: Multi-USIM Problematic Reachability 
& For MUSIM devices, active traffic on one SIM can prevent the other SIM from receiving paging, which an attacker can exploit to disrupt reachability
& U.S. carrier networks;\newline Samsung S24,\newline Samsung S21
& Repeated attacker calls to SIM1 prevented SIM2 from ringing across tested carrier combinations
& Specification and implementation gap for shared-transceiver MUSIM devices, where reliable paging monitoring for the inactive SIM is not guaranteed \\
\cline{2-6}

& V2: Emergency Session Teardown via Forged Signaling to Victim
& Unprotected emergency REGISTRATION and SIP signaling enable MitM injection that terminates emergency sessions
& srsRAN/srsEPC with Linphone IMS with a controlled emergency session setup
& Forged RRC RELEASE or SIP CANCEL terminated the victim's emergency session
& A threefold design weakness in emergency services: unprotected emergency NAS messages, acceptance of unauthenticated RRC RELEASE, and unprotected SIP signaling \\
\hline

TS 23.501\newline
v16.8.0~--\newline
v20.2.0$^\dagger$
& V3: Vulnerable DNS-Based PLMN Discovery over Non-3GPP Access
& In non-3GPP access, an attacker can use forged DNS responses to redirect a UE to malicious infrastructure before any security context is established
& Controlled Wi-Fi network; \newline Samsung S24,\newline Moto G Stylus 5G
& DNS poisoning blocked legitimate N3IWF/ePDG access and non-3GPP registration
& Unauthenticated DNS-based PLMN discovery over untrusted pre-attachment links enables network misdirection attacks \\
\hline

TS 24.229\newline
v16.8.0~--\newline
v20.0.0$^\dagger$
& V4: Privacy Exposure via Over-Disclosed VoIMS Signaling Headers
& Over-disclosed SIP and SDP headers expose device, user identity, and IMS infrastructure information during VoIMS signaling
& U.S. carrier networks; \newline Samsung S21, \newline Pixel 5
& SIP traces exposed private information
& The text-based SIP signaling design largely increases the risk of transmitting headers with private information \\
\hline

TS 24.301\newline
v16.8.0~--\newline
v20.0.0$^\dagger$ 
& V5: Exploitable CS-Fallback Initiation
& An attacker sniffs an unprotected extended service request and injects a forged SERVICE REJECT (with EMM cause 39 and T3442). It can trap the victim in a long throttling timer
& srsRAN/open5GS SDR testbed; \newline Samsung S24,\newline Moto G Stylus 5G
& Devices accepted forged SERVICE REJECT and stopped CSFB call setup until T3442 expired
& Specifications allow SERVICE REJECT to be processed without ciphering or integrity protection as a fail-safe mechanism, enabling attackers to inject unauthenticated rejection messages \\
\hline

\multirow[t]{2}{=}{\raggedright
TS 38.331\newline
v17.0.0~--\newline
v19.3.0$^\dagger$}
& V6: Access Trap via Access Barring Factor
& Rogue base station advertises restrictive barringFactor in SIB1/SIB2, causing UEs to suppress MO access attempts and refrain from initiating RRC connection establishment
& RF-shielded srsRAN, srsEPC testbed; \newline Samsung S24,\newline Moto G Stylus 5G
& Devices self-blocked access after receiving SIB2 with ac-BarringFactor set to 0
& UAC policies are broadcast via unauthenticated SIB messages before security is established, creating a pre-authentication trust gap \\
\cline{2-6}

& V7: Rogue Base Station RRC RESUME Blackhole
& By exploiting unauthenticated RRC RESUME procedure, a rogue gNB traps the victim UE in a continuous loop of connection failures
& RF-shielded gNB setup; \newline Samsung S21, \newline Moto G Stylus 5G
& Devices were attracted to the rogue gNB and repeatedly failed RRC RESUME
& Cell selection and reselection occur at the RRC layer before a protected AS channel is established, and the mechanism does not account for resume failures \\
\hline

\multicolumn{6}{|l|}{$\dagger$ The latest 3GPP specification version available at the time of submission.}  \\
\hline
\end{tabular}
\vspace{-0.11in}
\end{table*}

Among the well-studied TS~38.331, TS~24.501, and TS~24.301, \lm uncovers 5 new vulnerabilities.
Beyond these three specifications, we further apply \lm to TS~23.501 (System architecture for the 5G System (5GS)) and TS~24.229 (IP multimedia call control protocol based on Session Initiation Protocol (SIP) and Session Description Protocol (SDP); Stage 3) to demonstrate the extensibility of \lm beyond commonly studied NAS and RRC specifications. 
Considering that validating each finding requires substantial human effort, \lm analyzes only one selected section from TS~23.501 and TS~24.229, respectively; \lm identifies two additional vulnerabilities from selected sections. 
Table~\ref{tab:new_vulnerabilities} summarizes all 7 newly identified vulnerabilities with the precise description of the vulnerability, how we conduct the proof-of-concept validation, and the root cause analysis. All 7 findings were validated through controlled experiments using U.S. carrier networks or SDR-based testbeds.

\noindent 
\textbf{Not every violation flagged by \lm is ultimately reported as a confirmed vulnerability here for two reasons.}
(1) \ul{Some flagged issues cannot be experimentally examined due to the missing support in operational infrastructure. Nevertheless, some flagged issues reflect the same concerns raised by domain experts through 3GPP Change Requests (CRs) before the real-world deployment.}
For example, since Release~17, TS~23.501 introduces the Unavailability Period mechanism to help the network manage unreachable states of NTN (non-terrestrial networks) UEs. 
The specification defines how the AMF determines the loss-of-coverage schedule, including the Start of Unavailability Period and the Unavailability Period Duration. 
However, it does not fully specify the UE-side handling of these configurations, despite their direct impact on paging reception, access stratum operation, and power-saving behavior. 
This creates a potential design loophole: an NTN UE and the AMF may operate with inconsistent Unavailability Period settings, which can disrupt mobility procedures and trigger transient connectivity failures and unnecessary re-registration attempts. 
Unfortunately, we did not observe support for the Unavailability Period mechanism in either operational infrastructure or SDR-based platforms. 
Importantly, after reviewing CRs used by 3GPP to create revised versions of specifications, we found that this out-of-sync Unavailability Period issue was also raised by Samsung in CR~5614~\cite{netovate_cr_search}. 
However, the proposed change does not fully eliminate the inconsistency risk. 
This alignment with an independently reported CR confirms the value of \lm's findings, even when full experimental validation is not feasible. 
Consequently, we submit these potential issues to GSMA and 3GPP rather than reporting them as confirmed vulnerabilities. 
(2) \ul{Confirmed vulnerability reporting requires substantial human effort to investigate identified issues and run validation experiments.} 
Although \lm is extensible to a broad range of 3GPP standards, in this work, we focus on a subset of specifications to demonstrate \lm's effectiveness in discovering new vulnerabilities. 

We next present V1 and V2 in detail below as representative examples, including introductions of vulnerabilities, proof-of-concept validations, and root cause \& lessons learned discussions. The remaining V3--V7 vulnerabilities are elaborated in Appendix~\ref{append:new_vulnerabilities_v1_v7} following the same structure.

\subsubsection{V1: Multi-USIM problematic reachability}

\vspace{-0.1in}
\begin{figure}[H]
    \centering
    \includegraphics[page=1,width=0.8\linewidth]{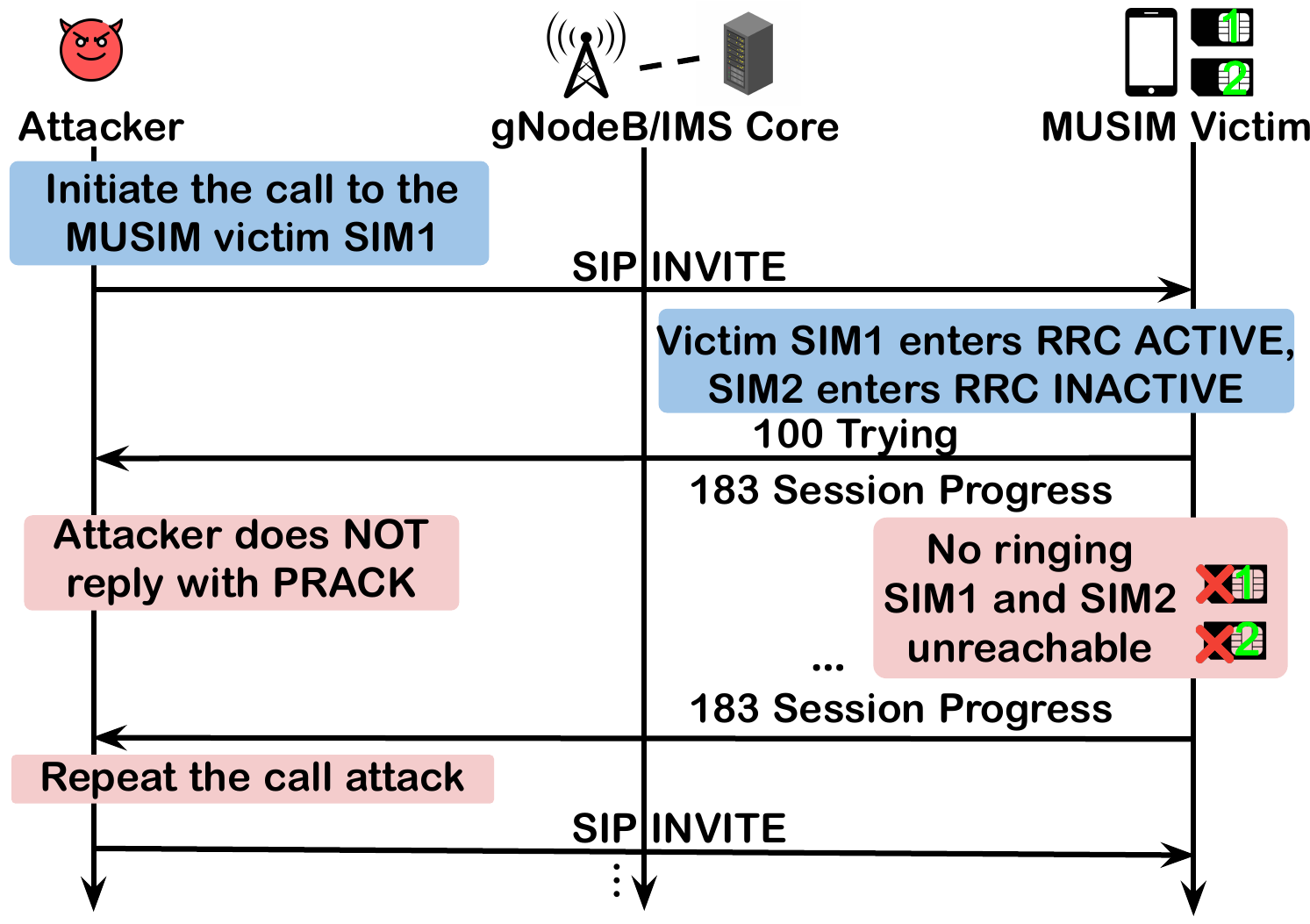}
    \vspace{-0.1in}
    \caption{Multi-USIM Problematic Reachability}
    \label{fig:musim-dos}
    \vspace{-0.1in}
\end{figure}

5G introduces the RRC\_INACTIVE state to balance power efficiency with low-latency responsiveness. In this state, a UE remains in 5GMM\_CONNECTED mode while its RRC connection is suspended and its session context is stored. This design allows the UE to avoid full reconnection procedures, thereby saving energy and reducing signaling overhead. The mechanism is particularly relevant for multi-USIM (MUSIM) devices, as it allows a SIM in the RRC\_INACTIVE state to quickly restore its session to respond to a paging request. This appears to be a reasonable approach, especially for addressing paging collisions~\cite{3gpp2022multisim} by enabling the SIM to quickly receive the message before the request times out.

However, in practice, most modern UEs (e.g., iPhone, Samsung Galaxy, Google Pixel) operate as Dual SIM Dual Standby, sharing a single RF transceiver. Under this hardware constraint, the design is not robust: when the transceiver is fully occupied by SIM1, SIM2 cannot monitor paging occasions or process paging requests. Despite the fact that paging includes a coarse paging cause (e.g., voice, data), the inactive SIM can still miss paging occasions and must rely on network re-paging, creating an exploitable reachability gap.

\noindent $\diamond$ {Proof-of-concept Attack.}  The victim is a MUSIM user with two SIMs on a commodity off-the-shelf (COTS) phone. 
The attacker aims to make both SIM subscriptions of the victim persistently unreachable. It assumes that at least one of the victim’s SIM numbers can be obtained from publicly available sources, such as social networking platforms, enterprise contact directories, or leaked customer databases. 
With this minimal knowledge, the attacker can launch a stealthy DoS attack against both SIMs.  

\vspace{-0.1in}
\begin{table}[H]
\centering
\caption{Multi-USIM DoS attack results}
\vspace{-0.1in}
\label{tab:dual_sim_ringing_data}
\small
\begin{tabular}{l l l c}
\hline
\textbf{Phone} & \textbf{SIM1} & \textbf{SIM2} & \textbf{Ringing?} \\
\hline
Samsung S24 & OP-I & OP-II  & x \\
Samsung S24 & OP-I & OP-I   & x \\
Samsung S24 & OP-I & OP-III & x \\
Samsung S21 & OP-I & OP-II  & x \\
Samsung S21 & OP-I & OP-I   & x \\
Samsung S21 & OP-I & OP-III & x \\
\hline
\end{tabular}
\end{table}

As illustrated in Figure~\ref{fig:musim-dos}, the attack proceeds as follows.
The attacker first initiates a call to SIM1, which establishes an active RRC connection for SIM1 and forces SIM2 into RRC INACTIVE. To avoid alerting the victim, the attacker can leverage the ghost call attack~\cite{lu2020ghost,wang2021insecurity}, preventing the call from ringing while keeping SIM1 trapped in the RRC ACTIVE state during the entire attack period. 
Note that, under the ghost call attack, SIM1 is also unable to receive any other incoming calls. Consequently, SIM2 remains stuck in RRC INACTIVE. All incoming paging requests for SIM2 will be ignored. The device neither rings nor displays an incoming call notification. As a result, the victim UE becomes persistently unreachable through both SIM1 and SIM2, despite having two independent subscriptions.

We validate this attack on a Samsung Galaxy S24 and a Samsung Galaxy S21. We choose these devices because both provide two physical SIM slots, which simplifies repeated experiments and SIM swapping without triggering the additional security restrictions often associated with moving eSIM profiles across devices. During the interval in which the attacker withholds \textsf{PRACK}, we use another phone to call SIM2. As summarized in Table~\ref{tab:dual_sim_ringing_data}, neither device produces a ringing notification for SIM2 on any tested carrier. Notably, this attack does not strictly require a ghost call. The attacker can also keep repeatedly calling SIM1 or maintain an ongoing connection with SIM1 to occupy the shared RF resources. Although this variant is more likely to attract the victim’s attention, SIM2 still does not receive incoming call notifications during the attack period.

\noindent $\diamond$ \textbf{Root Cause and lessons learned.} 
The root cause is a mismatch between specification assumptions and widely deployed MUSIM hardware. The 5G NAS specification~\cite{3GPP_TS_24501} implicitly assumes device capabilities that can maintain simultaneous accessibility for multiple SIMs, whereas most modern MUSIM devices share a single RF module. The specifications do not mandate mechanisms that ensure all MUSIM devices, in particular, devices with a shared transceiver, can reliably monitor and process Paging Requests. This gap enables an attacker to exploit the MUSIM paging mechanism to attack MUSIM reachability.  Mitigating this issue likely requires combined core-network and device-side enhancements in future 3GPP releases. For example, MUSIM devices could implement adaptive internal scheduling for ongoing traffic/calls and bring SIMs in RRC INACTIVE to ACTIVE for paging monitoring, rather than ignoring paging on the inactive SIM. 

\vspace{-0.10in}
\subsubsection{V2: Emergency Session Teardown via Forged Signaling to Victim} 
\noindent To ensure emergency services remain accessible to all users, 5G NAS specification TS 24.501~\cite{3GPP_TS_24501} permits UEs without valid credentials to connect. If an unregistered UE attempts to make an emergency call (e.g., 911 call), it initiates a NAS REGISTRATION REQUEST with the type set to emergency registration. Since this message may be transmitted without encryption or an established NAS security context, it opens a loophole that adversaries can monitor and use a rogue base station to send fabricated responses to the UE, enabling a man-in-the-middle (MiTM) attack. 
The exposure extends to the IMS layer: TS~24.229 permits emergency SIP signaling without authentication or IPsec protection, leaving critical emergency call-control messages vulnerable to interception and manipulation~\cite{hu2024uncovering, 10479537, inproceedings,10.1145/3599184.3599195}. 

\noindent \textbf{Is it the same vulnerability in~\cite{hu2022uncovering}?} This vulnerability is fundamentally different from previously reported vulnerabilities~\cite{hu2022uncovering}, which exploit the fact that the infrastructure accepts unverifiable emergency session requests over the IP-CAN session and suffers from improper cross-layer security binding. These are infrastructure-side logic loopholes, and operators could mitigate them by temporarily allowing duplicate emergency IP-CAN sessions. Our V2 targets a loophole in the standard-permitted emergency entry path. It enables a MiTM attack at the radio and emergency signaling layers directly targeting the UE; therefore, V2 cannot be noticed by the carrier networks.

\begin{figure}[t]
    \centering
    \includegraphics[page=2,width=0.8\linewidth]{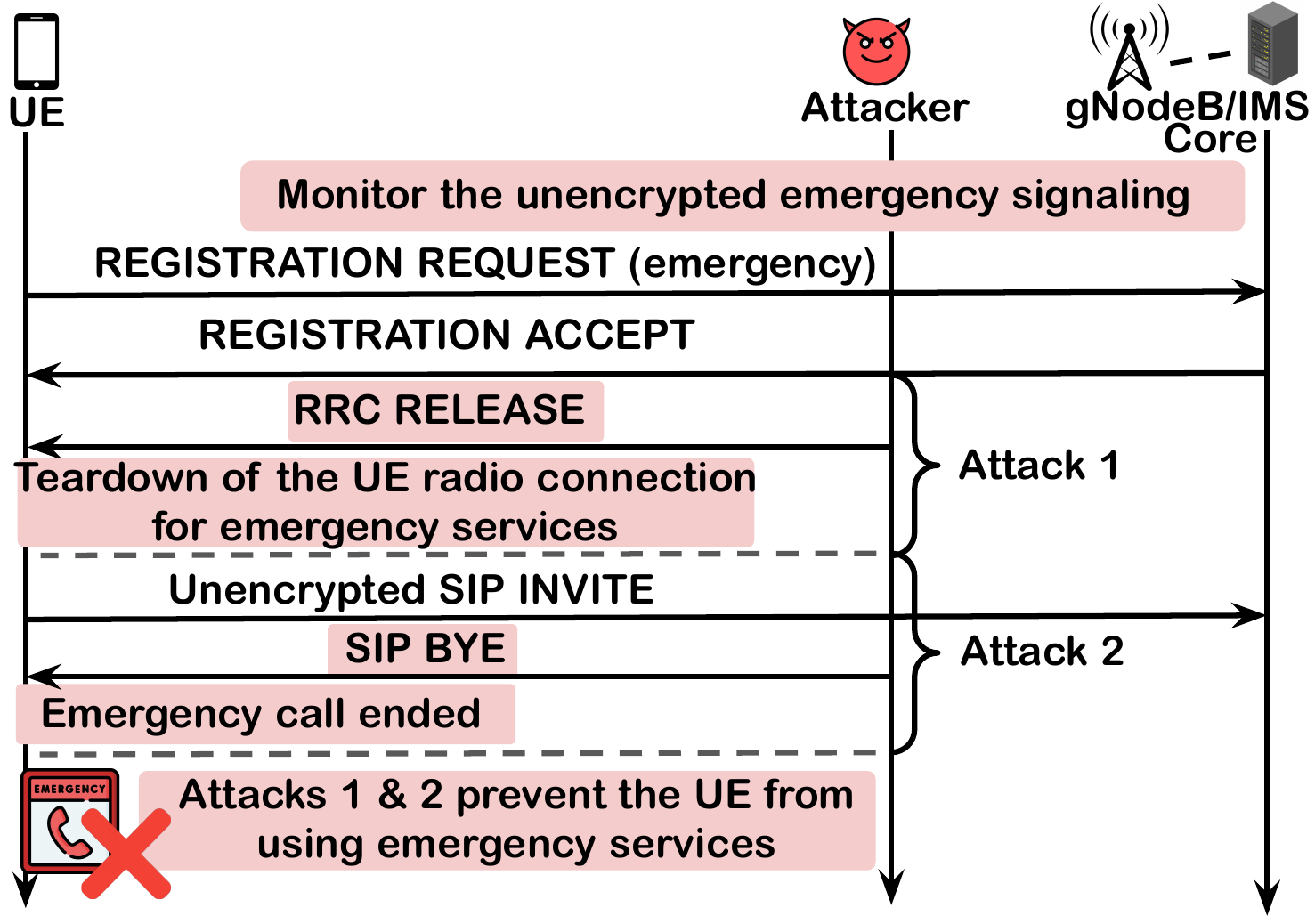}
    \vspace{-0.15in}
    \caption{Emergency Session Teardown via Forged Signaling to Victim}
    \label{fig:emergency}
    \vspace{-0.21in}
\end{figure}

\noindent $\diamond$ \textbf{Proof-of-concept Attacks.}
Two proof-of-concept attacks are used to validate this vulnerability. 

\noindent \textbf{Attack 1: Forcing emergency radio session release.} 
This attack exploits the NAS registration and RRC release procedures. When an unregistered UE initiates an emergency call, it sends a NAS REGISTRATION REQUEST with the type set to emergency. Because the UE does not have a valid security context, this message is transmitted unprotected. Therefore, an attacker operating a rogue base station can capture both REGISTRATION REQUEST and the REGISTRATION ACCEPT from the legitimate network. Next, the attacker can send a forged RRC RELEASE to the UE, forcing it to tear down the ongoing radio connection for the upper emergency services. 

Due to ethical considerations, we validate this attack in a fully controlled environment rather than interacting with the operational networks or real 911 call center. Specifically, we deploy a serving network using srsRAN and srsEPC~\cite{srsran}, together with an IMS service based on the Linphone VoIP server~\cite{linphone}. Additionally, we set up another set of srsRAN and srsEPC as the attacker on the same server. We use UERANSIM~\cite{ueransim} as the victim UE. 
After the UE receives the Registration Accept, the attacker keeps sending a forged RRC RELEASE to the victim UE. In wild settings, the attacker needs to do an RF scan, detect SSB/PBCH, and read MIB/SIB1 once synchronized for sending the forged RRC message on the same downlink carrier the victim is using. 
Figure~\ref{fig:emergency} shows the victim attempts to reconnect to continue the emergency service.   
The attacker can continue sending the forged RRC RELEASE to prevent the victim from getting the radio connection.

\begin{figure}[t]
    \centering
    \includegraphics[page=2,width=0.95\linewidth]{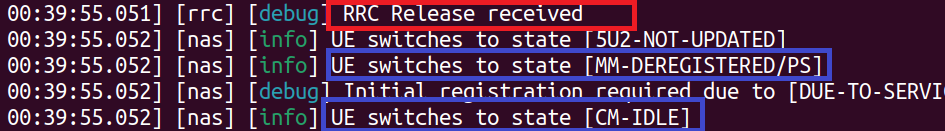}
    \vspace{-.1in}
    \caption{The forged RRC Release message traps the victim UE by disconnecting it from the base station}
    \label{fig:emergency}
    \vspace{-0.25in}
\end{figure}

\noindent \textbf{Attack 2: Ghost emergency call cancellation.} 
It targets the unprotected SIP signaling for emergency voice services. In this attack, the attacker leverages the fact that the 3GPP specification permits emergency services without authentication or encryption. The attacker intercepts the unencrypted SIP INVITE and spoofs a forged SIP CANCEL message, deceiving the emergency UE that the emergency call session has been legitimately terminated. We validate this attack using the aforementioned in-house setting for ethical considerations. After observing the unencrypted SIP INVITE, the attacker can send a forged SIP CANCEL to the victim to cancel the call. Our result confirms that the victim's emergency call is canceled after receiving a `ghost' cancellation signaling message.   

\noindent $\diamond$ \textbf{Root Cause and Lesson Learned.} 
These attacks stem from the lack of protection in the REGISTRATION REQUEST and the design trade-offs in the emergency service procedures to guarantee universal accessibility. Together, they expose unprotected NAS and SIP signaling to attackers, making mobile users vulnerable to emergency service DoS attacks. 
The root cause is therefore threefold: (1) the omission of mandatory encryption and authentication for emergency NAS messages, (2) the acceptance of RRC RELEASE without protections, and (3) the allowance of unprotected SIP signaling for emergency voice services. Future 3GPP specification releases must revisit the emergency service procedures and incorporate mitigations that preserve accessibility without sacrificing security. For instance, lightweight integrity-protection mechanisms could be mandated for emergency signaling, ensuring messages cannot be silently modified. Different from~\cite{hu2022uncovering}, our proposed attacks target emergency UEs; therefore, they are harder to be noticed by carriers.

\subsection{RQ2: Vulnerability Discovery Comparison}
Beyond discovering new vulnerabilities, it is important to understand how \lm performs relative to prior art. 
Accordingly, we compare \lm with the state-of-the-art approach in terms of vulnerability discovery effectiveness, which is intended to assess whether \lm can have a better coverage of security flaw identification in 3GPP specifications. As discussed in Section~\ref{sect:related}, some existing approaches~\cite{chen2021bookworm,chen2023sherlock,tan2025automated} target implementation-level issues instead of detecting design flaws in 3GPP specifications. \textit{CellularLint}~\cite{rahman2024cellularlint} aims to detect sentence-level semantic conflicts in 3GPP specifications rather than finding the design vulnerabilities.  
To the best of our knowledge, \textit{Hermes}~\cite{al2024hermes} is the only comparable system that can also discover vulnerabilities from standards. To enable a fair comparison, we collect and develop a benchmark set of known vulnerabilities~\cite{shaik2015practical, 6550445, kambourakis2011attacks, lee2009detection, leong2014unveiling, kim2019touching, van2015defeating, park2016white, chlosta2019lte, 8958725, michau2016not, hussain20195greasoner, borgaonkar2018new, 8894379, chlosta20215g, hussain2019privacy, al2024hermes, hussain2018lteinspector} that are design defects from same specifications (TS~38.331, TS~24.501, TS~24.301) used by \textit{Hermes}.

As shown in Table~\ref{tab:vulnerabilities}, \textit{Hermes} can detect 22 of all 36 known vulnerabilities in 5G and 4G NAS and RRC specifications. 
After manually verifying the candidate violations flagged by \lm, we confirm that flagged violations can cover all of these KNOWN vulnerabilities.
The vulnerability discovery results show that \lm is capable of effectively modeling mobile network procedures, conducting security analysis, and identifying security threats in 3GPP specifications. Combined with the results in RQ1, \lm not only has better vulnerability discovery coverage, but also can discover new vulnerabilities. 

\noindent \textbf{Why are there more known vulnerabilities than new vulnerabilities?}
The 36 known vulnerabilities form a cumulative benchmark collected from the aforementioned prior studies over several years, whereas the 7 new vulnerabilities include only previously unreported, experimentally validated design flaws identified in this work. As clarified in RQ1, we also exclude duplicate, implementation-dependent, and unconfirmed candidates from the new-Reference Guided Connection count.

\begin{table}[ht]
\centering
\caption{Comparison of KNOWN vulnerabilities detected by Hermes and \lm}
\vspace{-0.1in}
\label{tab:vulnerabilities}
\scriptsize
\begin{tabularx}{\linewidth}{p{.1cm} p{5.7cm} c c}
\toprule
\textbf{ID} & \textbf{Attack} & \textbf{H*} & \textbf{C**} \\
\midrule
1 & Downgrade to non-LTE network services \cite{shaik2015practical} & \cmark & \cmark \\
2 & Denying all network services \cite{shaik2015practical} & \cmark & \cmark \\
3 & Denying selected service \cite{shaik2015practical} & \xmark & \cmark \\
4 & Signaling DoS \cite{6550445, kambourakis2011attacks, lee2009detection, leong2014unveiling} & \cmark & \cmark \\
5 & S-TMSI catching \cite{kim2019touching}& \cmark & \cmark \\
6 & IMSI catching\cite{van2015defeating}& \cmark & \cmark \\
7 & EMM Information \cite{park2016white}& \cmark & \cmark \\
8 & Impersonation attack \cite{chlosta2019lte} & \xmark & \cmark \\
9 & Synchronization Failure attack \cite{8958725} & \xmark & \cmark \\
10 & Malformed Identity Request \cite{michau2016not} & \xmark & \cmark \\
11 & Neutralizing TMSI refreshment \cite{hussain20195greasoner} & \xmark & \cmark \\
12 & NAS Counter Reset \cite{hussain20195greasoner} & \cmark & \cmark \\
13 & Uplink NAS Counter Desynchronization \cite{hussain20195greasoner} & \cmark & \cmark \\
14 & Exposing NAS Sequence Number \cite{hussain20195greasoner} & \cmark & \cmark \\
15 & Cutting off the Device \cite{hussain20195greasoner} & \xmark & \cmark \\
16 & Exposure of SQN \cite{borgaonkar2018new} & \cmark & \cmark \\
17 & 5G AKA DoS Attack\cite{8894379} & \cmark & \cmark \\
18 & SUCI catching \cite{chlosta20215g} & \xmark & \cmark \\
19 & IMSI cracking \cite{hussain2019privacy} & \xmark & \cmark \\
20 & NAS COUNT update attack \cite{al2024hermes} & \cmark & \cmark \\
21 & Deletion of allowed CAG list \cite{al2024hermes} & \cmark & \cmark \\
22 & Downgrade using ATTACH/REGISTRATION REJECT \cite{shaik2015practical} & \cmark & \cmark \\
23 & AUTHENTICATION REJECT attack \cite{8958725} & \cmark & \cmark \\
24 & DETACH/DEREGISTRATION REQUEST attack \cite{hussain2018lteinspector} & \cmark & \cmark \\
25 & SERVICE REJECT attack \cite{shaik2015practical} & \cmark & \cmark \\
26 & Denial-of-Service with RRC SETUP REQUEST attack \cite{hussain20195greasoner} & \xmark & \cmark \\
27 & Installing Null Cipher and Null Integrity \cite{hussain20195greasoner} & \cmark & \cmark \\
28 & Lullaby Attack \cite{hussain20195greasoner} & \cmark & \cmark \\
29 & Incarceration with RRC REJECT/RELEASE \cite{hussain20195greasoner} & \cmark & \cmark \\
30 & Measurement report \cite{shaik2015practical} & \xmark & \cmark \\
31 & RLF report \cite{shaik2015practical} & \cmark & \cmark \\
32 & Blind DoS attack \cite{kim2019touching} & \xmark & \cmark \\
33 & AKA bypass \cite{kim2019touching} & \xmark & \cmark \\
34 & Paging channel hijacking \cite{hussain2018lteinspector} & \xmark & \cmark \\
35 & Energy Depletion with RRC SETUP \cite{al2024hermes} & \cmark & \cmark \\
36 & V2X Message Spoofing over PC5 \cite{sedar2023comprehensive} & \xmark & \cmark \\
\midrule
\multicolumn{2}{l}{\textbf{Detected Ratio}} & \textbf{22/36} & \textbf{36/36} \\
\multicolumn{4}{l}{H*: \textit{Hermes}, C**: \lm} \\
\bottomrule
\end{tabularx}
\vspace{-0.2in}
\end{table}

\subsection{RQ3: How effectively does SCA preserve procedural information?}

SCA nodes are the core modeling representation of \lm. As introduced in Section~\ref{subsect:sca_representation}, each SCA node represents one self-contained transition using four fields. The vulnerability-discovery results in RQ1 and RQ2 demonstrate the effectiveness of \lm, but do not explain whether SCA preserves the procedural information (i.e., states, conditions, and actions) required for downstream function-chain construction and security analysis in \textit{Function Chain Builder} and \textit{SecOracle}. Because FSMs are the dominant representation used by prior specification-analysis systems, we next compare SCA nodes with the conventional FSMs synthesized by prior works. The goal of this evaluation is to isolate the contribution of SCA and examine whether SCA can better preserve the procedural information needed for downstream tasks.

\textit{Hermes}~\cite{al2024hermes} is the closest reference system, which extracts protocol models from 3GPP specifications for security analysis. \textit{ARCANE}~\cite{tan2025automated} is also included, while its FSMs are designed for impleme-\ ntation-level model-based fuzzing rather than specification-level design-flaw analysis. All three approaches are applied to the same 5G/4G NAS and RRC specifications. Because conventional FSMs do not expose conditions and actions as explicit fields, we normalize the representations into the same four-field structure: start state, condition, action, and end state. For \textit{Hermes}, condition and action information is derived from its parsed FSM transitions. For \textit{ARCANE}, we adopt its manual approach to map its message-level transitions and information-element constraints into the same structure. We then showcase the evaluation using three metrics: quantity, completeness, and accuracy.

\begin{figure}[t]
    \centering
    \includegraphics[width=1\linewidth]{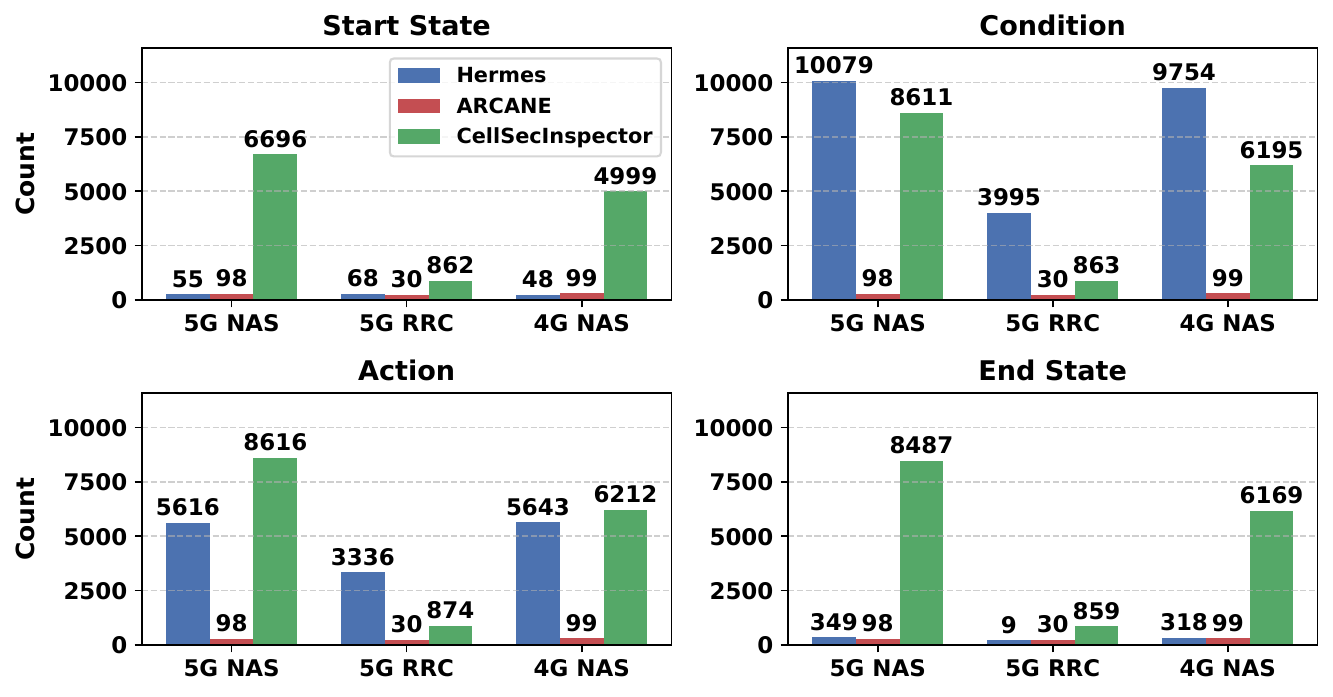}
    \vspace{-.3in}
    \caption{Quantity Comparison of SCA nodes and FSMs}
    \label{fig:sca_statistics}
    \vspace{-0.1in}
\end{figure}

\noindent \textbf{Quantity.} Figure~\ref{fig:sca_statistics} shows the number of extracted fields across the three specifications. There are two key observations. (1) SCA captures substantially more start and end states than the two FSM-based approaches, indicating more derived mobile network functions from 3GPP specifications. Table~\ref{tab:sca_vs_hermes} compares an SCA node derived by \lm and an FSM transition annotated by \textit{Hermes} for a sentence `\textit{A UE enters the state 5GMM-SERVICE-REQUEST-INITIATED after it has started the service request procedure and is waiting for a response from the network.}' from TS 24.501. While the sentence does not explicitly specify the start state, \lm reasonably infers that the UE is likely in the 5GMM-REGISTERED state and correctly recognizes two sequential conditions, the action, and the end state. (2) \textit{Hermes} reports a larger number of extracted conditions and actions in 5G and 4G NAS and RRC specifications. However, it is primarily due to splitting closely connected conditions and actions within a single sentence into multiple independent entries, as shown in Table~\ref{tab:sca_vs_hermes}. This fragmentation inflates raw counts of extracted elements but fails to preserve the internal structure and logical interdependencies among conditions and actions.

\begin{table}[t]
\centering
\caption{Comparison between an SCA node from \lm and an FSM transition from \textit{Hermes}}
\vspace{-0.1in}
\label{tab:sca_vs_hermes}
\scriptsize
\setlength{\tabcolsep}{2.8pt}
\begin{tabular}{p{0.9cm} | p{3.1cm} | p{3.9cm}}
\toprule
\textbf{Element} & \textbf{\lm (SCA Node)} & \textbf{\textit{Hermes} (FSM Transition)} \\
\midrule
\textbf{Start State} &
UE is likely in 5GMM-REGISTERED. &
\texttt{<start\_state>} A UE enters the state 5GMM-SERVICE-REQUEST-INITIATED \texttt{</start\_state>} \\
\midrule
\textbf{Condi- tion} &
After the UE has started the service request procedure AND is waiting for a response from the network. &
\texttt{<condition>} it has started the service request procedure \texttt{</condition>} \newline
\texttt{<condition>} is waiting for a response from the network \texttt{</condition>} \\
\midrule
\textbf{Action} &
UE enters the 5GMM-SERVICE-REQUEST-INITIATED state. &
N/A \\
\midrule
\textbf{End State} &
UE is in the 5GMM-SERVICE-REQUEST-INITIATED state, waiting for a response from the network. &
N/A \\
\bottomrule
\end{tabular}
\end{table}

\begin{table}[t]
\centering
\caption{
SCA completeness comparison of Hermes, ARCANE, and \lm
}
\vspace{-0.1in}
\label{tab:sca_completeness}
\scriptsize
\setlength{\tabcolsep}{2.2pt}
\begin{tabular}{p{0.5cm}lccccc}
\toprule
\textbf{Spec\#} & \textbf{Method} & \textbf{0 fields} & \textbf{1 field} & \textbf{2 fields} & \textbf{3 fields} & \textbf{4 fields} \\
\midrule

\multirow{3}{*}{24.501}
 & Hermes  & 1122 (20.1\%) & 1055 (19.0\%) & 2177 (39.1\%) & 207 (3.7\%) & 5 (0.1\%)   \\
 & CellSec & 1778 (17.1\%) & 36 (0.3\%)    & 132 (1.3\%)  & 1920 (18.4\%) & 6549 (62.9\%) \\
 & ARCANE  & 0 (0\%) & 0 (0\%) & 0 (0\%) & 0 (0\%) & 98 (100\%) \\

\midrule

\multirow{3}{*}{38.331}
 & Hermes  & 2948 (46.2\%) & 2675 (41.9\%) & 753 (11.8\%) & 1 (0.0\%) & 0 (0.0\%)   \\
 & CellSec & 321 (26.9\%)  & 11 (0.9\%)    & 1 (0.1\%)   & 7 (0.6\%) & 855 (71.5\%) \\
 & ARCANE  & 0 (0\%) & 0 (0\%) & 0 (0\%) & 0 (0\%) & 30 (100\%) \\

\midrule

\multirow{3}{*}{24.301}
 & Hermes  & 3026  (50.6\%) & 463 (7.7\%)   & 2244 (37.5\%) & 237 (4.0\%) & 6 (0.1\%)   \\
 & CellSec & 1771 (22.2\%) & 30 (0.4\%)    & 13 (0.2\%)   & 1207 (15.1\%) & 4974 (62.2\%) \\
 & ARCANE  & 0 (0\%) & 0 (0\%) & 0 (0\%) & 0 (0\%) & 99 (100\%) \\

\bottomrule
\end{tabular}
\vspace{-0.15in}
\end{table}

\noindent \textbf{Completeness.} We evaluate the completeness of state transitions. A state transition is defined to be complete if all 4 corresponding start, condition, action, and end fields can be derived and are not placeholders (e.g., ``Not specified'' or ``Not explicitly defined''). More complete fields indicate that more information is likely derived from 3GPP specifications. 

\begin{table}[t]
\centering
\caption{Expert evaluation results}
\vspace{-0.1in}
\label{tab:expert_evaluation}
\scriptsize
\setlength{\tabcolsep}{5pt}
\renewcommand{\arraystretch}{1.00}
\begin{tabular}{lccccc}
\multicolumn{6}{c}{\textbf{(a) Expert judgment of SCA nodes and FSMs}} \\
\midrule
\textbf{Method} & \textbf{Total} & \textbf{Accuracy} & \textbf{$\{a,b,c,d\}$} & \textbf{Raw Agr.} & \textbf{$\kappa$} \\
\midrule
\textit{CellSecInspector} & 3147 & 98.92\% & $\{3113,13,8,13\}$ & 99.33\% & 0.80 \\
\textit{Hermes} & 3634 & 25.23\% & $\{917,32,28,2657\}$ & 98.35\% & 0.96 \\
\textit{ARCANE} & 162 & 0.00\% & $\{0,0,0,162\}$ & 100.00\% & N/A \\
\midrule
\addlinespace[2.2pt]

\multicolumn{6}{c}{\textbf{(b) Expert validation of SCA node extraction across 4 models}} \\
\midrule
\textbf{Model} & \textbf{Expert1} & \textbf{Expert2} & \textbf{Avg. Acc.} & \textbf{Cohen's $\kappa$} & \\
\midrule
DeepSeek-V3.2 & 1.00 & 1.00 & 1.00 & 1.00 & \\
Qwen-Plus & 1.00 & 1.00 & 1.00 & 1.00 & \\
Qwen3-4B & 0.90 & 0.90 & 0.90 & 0.92 & \\
DeepSeek-R1-8B & 0.90 & 1.00 & 0.95 & 0.94 & \\
\midrule
\end{tabular}
\vspace{-0.26in}
\end{table}

As shown in Table~\ref{tab:sca_completeness}, it is observed that a large portion of SCA transitions contain three or four valid fields, and 62.2\%--71.5\% of the extracted transitions are fully specified across the three specifications. \textit{Hermes} extracts many partial transitions, while \textit{ARCANE} produces a much smaller set of structurally complete message-level transitions because its representation is intentionally tailored to fuzzing. These results show that SCA provides a useful balance between broad procedural coverage and preservation of the information required for downstream security analysis.

\noindent \textbf{Accuracy.} We evaluate whether an SCA node or an FSM is accurate based on two criteria: (1) evidence-grounded correctness, whether each extracted representation accurately reflects the protocol behavior described in the specification and is supported by explicit textual evidence, and (2) semantic consistency, whether the extracted representation preserves the intended meaning of the mobile network function behavior described in 3GPP specifications. Because this judgment requires domain expertise in cellular network standards, it cannot be reliably outsourced to crowdworkers. We thus ask two domain experts to independently judge whether an SCA node/an FSM is accurate. Formally, each SCA node and FSM is labeled as $v_i \in \{Pass, Fail\}$. We define that an SCA node or FSM is accurate only if both experts label it as Pass. Based on the annotations from two experts, we construct a contingency table $\{a,b,c,d\}$, where $a$ denotes Pass/Pass, $b$ denotes Pass/Fail, $c$ denotes Fail/Pass, and $d$ denotes Fail/Fail. Due to the large amount of SCA nodes and FSMs, two experts are asked to assess the SCA nodes and FSMs extracted from the selected sections, including Clause 4 of 4G NAS specification, Clause 4 of 5G NAS specification, Clauses 4 and 5 of 5G RRC specification. 
We report the acceptance rate ($a/N$), raw agreement ($(a+d)/N$), and Cohen's $\kappa$ in Table~\ref{tab:expert_evaluation} (a).

The results show that the SCA representation more consistently preserves 3GPP specification-grounded and semantically coherent protocol behaviors. The main advantage is not the use of a different label format, but the explicit binding of the start state, triggering condition, prescribed action, and end state within one transition. By comparison, FSMs from \textit{Hermes} and \textit{ARCANE} may be suitable for their own downstream tasks (i.e., parsing or fuzzing). However, both experts agree that FSMs from \textit{Hermes} and \textit{ARCANE} struggle to model the mobile network functions.

We also assess whether the accuracy of extracted SCA nodes can be affected if \core is built on different vanilla models. Four models, \textbf{Qwen-Plus-2025-12-01}, \textbf{Qwen3-4B-Instruct-2507}, \textbf{DeepSeek-V3.2 Release 2025/12/01}, and \textbf{DeepSeek-R1-Distill-Llama-8B}, are used to adapt cellular domain knowledge and extract SCA nodes. As shown in Table~\ref{tab:expert_evaluation}(b), all four models achieve high sampled accuracy, while the larger models perform best. This result indicates that the SCA formulation is transferable across different base models and larger models are preferable choices since the capability still affects extraction quality.

\subsection{RQ4: Candidate Violation to Confirmed Vulnerability Conversion Rate}

As introduced in Sections~\ref{subsect:secoracle} and \ref{subsect:rq1_new_vulnerabilities}, not every candidate violation can be assessed and validated as a confirmed vulnerability. We therefore adopt strict, conservative criteria to measure the conversion rate from candidate violations to confirmed vulnerabilities: a candidate violation identified by \lm is classified as a confirmed vulnerability only if it satisfies the following criteria: (i) it is attributable to a standard-defined procedure; (ii) it is non-duplicate; (iii) it is not solely implementation-dependent; and (iv) it is experimentally validated when the required testing support is available. Candidates that are duplicates, implementation-dependent, or currently unverifiable due to missing operational or SDR support are not counted as confirmed vulnerabilities. 
Under these strict conservative criterion, \lm reports 90 candidate violations in TS~24.501, of which 18 are confirmed as valid design-defect instances; 62 candidates in TS~24.301, of which 15 are confirmed; 161 candidates in TS~38.331, of which 12 are confirmed; 36 candidates in the analyzed section of TS~23.501, of which 1 is confirmed; and 8 candidates in the analyzed section of TS~24.229, of which 1 is confirmed. Known vulnerabilities 22--25 in Table~\ref{tab:vulnerabilities} occur independently in both 5G and 4G specifications and are therefore counted as separate confirmed instances.

The relatively low conversion rate is mainly caused by repeated reports of the same root cause, implementation-dependent cases, and candidates that cannot yet be validated due to limited infrastructure support. For example, identical RRC messages reused across multiple TS~38.331 procedures can generate several candidate violations corresponding to one underlying weakness. Thus, these conversion rates should not be interpreted as the accuracy of the final confirmed vulnerability set. 
Instead, they are lower-bound rates over all candidate violations produced by \lm, which narrows a large specification analysis space to a smaller set of security-relevant candidates for expert review and validation.

\subsection{RQ5: Runtime Performance}

We measure the end-to-end processing time of key components, including \textit{SCA Representation Extractor}, \textit{Function Chain Builder}, \textit{SecOracle}, and \textit{VulnTestGenerator}. The reported runtime is measured using DeepSeek-V3.2~\cite{deepseek_v32_release_2025} via its public API through a workstation running Ubuntu 24.04 LTS with an NVIDIA RTX 4090 GPU.   
For the full specifications TS~24.501 (5G NAS), TS~24.301 (4G NAS), and TS~38.331 (5G RRC), and the selected sections of TS~23.501 (5G system architecture) and TS~24.229 (IMS signaling), the \textit{SCA Representation Extractor} derives 10,415, 7,995, 1,195, 1,253, and 320 SCA nodes in 17.36, 13.33, 1.99, 2.09, and 0.53 hours, respectively. The \textit{Function Chain Builder} is the primary runtime bottleneck. 
It takes approximately 60 days, 30 days, 4 days, 0.44 days, and 0.03 days to construct 11,273, 6,943, 222, 83, and 6 function chains, respectively, with an average of 5.46 seconds per analyzed SCA pair. Approximately 99\% of this time is spent identifying semantic and causal connections. Reference Guided Connection substantially reduces the candidate search space. \textit{Without it, exhaustive all-pair analysis becomes infeasible in our experimental setting. For example, on TS~24.501, our design reduces the estimated runtime from approximately 18.78 years ($10,415^2 \times \frac{5.46~\text{sec}}{3600 \times 24 \times 365} = 18.78 ~\text{years}$) to 60 days.} The \textit{SecOracle} and \textit{VulnTestGenerator} together complete vulnerability analysis and test case generation in 17.54, 10.80, 0.35, 2.10, and 0.54 hours for the five specifications, respectively. 

\noindent\textbf{Scalability limit.}
Although \lm improves the scalability of specification analysis, corpus-wide 3GPP analysis remains computationally challenging. Directly linking all SCA nodes across hundreds of specifications would create a prohibitively large pairwise search space, and the resulting semantic and causal checks would dominate runtime. Therefore, corpus-wide analysis requires further efforts, such as our reference-guided design that reduces the process from years to days, to achieve with limited computational resources. A practical deployment should prioritize security-critical specifications, such as the TS~23/24 control-plane families; \lm should be executed in parallel to boost the runtime performance. Overall, the current results demonstrate the feasibility of automated security analysis for large, security-critical specifications while identifying scalability as an important direction for future work.

\vspace{-0.08in}
\section{Discussion and Limitations}
\label{append:discussion}

\noindent \textbf{Can the defined security properties and four attacks guarantee discovery of all vulnerabilities?} 
\lm aims to reduce reliance on slow, manual expert review by enabling systematic and scalable detection of standard-level defects in the extracted function chains. It does not guarantee the discovery of all vulnerabilities for two main reasons.
First, security verification is inherently scoped. Any finite set of security properties may miss emerging or unmodeled failure modes. We can only claim security with respect to the predefined properties and assumptions of the threat model~\cite{chen2023sherlock,al2024hermes}. Second, while not all SCA nodes correspond to explicit message exchanges, some nodes represent event or condition-driven transitions such as local policy decisions. In this case, \textit{SecOracle} still applies the four attacks to such nodes by exploiting their triggered events or conditions. If an SCA node is purely deterministic without any inputs, such as entering one state after another state, this transition is non-attackable under our settings. Notably, the majority of reported vulnerabilities in RRC and NAS layers included in Table~\ref{tab:vulnerabilities} stem from message exchanges and thus fall within the scope of our attack model.   

\noindent  \textbf{Out-of-scope vulnerabilities.}
\lm currently targets standard design flaws that can be exposed through adversarial manipulation of message or event flows, including dropping, modifying, injecting, and replaying messages. This scope covers availability, authentication, authorization, integrity, confidentiality, and state-consistency issues in control-plane procedures. We admit that several important vulnerability classes are outside the current model. For example, \lm does not directly analyze cryptographic weaknesses, side channels, physical-layer or social-engineering attacks, and vendor deviations absent from specifications, and may miss deployment-specific vulnerabilities involving carrier policies or undocumented hardware.

\noindent\textbf{Specification evolution and cross-release analysis.} 

3GPP specifications continuously evolve across releases. Previously identified security issues may persist, be partially mitigated, or be eliminated by later revisions. These ongoing efforts are essential for strengthening the security of cellular systems. At the same time, the scale and complexity of the standards make it challenging to identify and eliminate every security-relevant design weakness through traditional methods. Particularly, we examined the specification text corresponding to our newly reported vulnerabilities in the latest available versions at the time of submission and found that all 7 vulnerabilities remain present, as shown in Table~\ref{tab:new_vulnerabilities}. This observation motivates the automated cross-release analysis. In future work, we plan to extend \lm to compare SCA nodes and function chains across versions, identify security-relevant changes, and determine whether previously identified weaknesses are mitigated, persist, or are reintroduced as specifications evolve.

\noindent\textbf{Ethical Consideration and Responsible Disclosure.}
Our research follows established ethical standards for security research. Identified vulnerabilities in cellular specifications are handled throu-\ gh responsible disclosure.
We work with stakeholders, including standardization organizations, operators, and device manufacturers, to report design defects. Their acknowledgments will be posted on our project website.
Our objective is not to expose flaws for exploitation, but rather to advance an efficient methodology for enabling 3GPP, operators, and vendors to rigorously review specifications prior to implementation and public deployment of services.

\section{Conclusion}
\noindent Security assurance in cellular networks relies on rigorous validation of 3GPP specifications, yet automation remains challenging due to complex semantics, cross-clause dependencies, and evolving standards. In this work, we present \lm, an automated framework for security assessment of 3GPP specifications. Through five coupled components, \lm transforms cellular network standards into structured SCA nodes, connects scattered procedures into function chains, checks them against security properties under four adversarial scenarios, and generates test cases for validation. Applying \lm to selected 5G and 4G specifications, we identify 43 vulnerabilities, including 36 known vulnerabilities and 7 new vulnerabilities. These results show that \lm can narrow large specifications to a small set of security-relevant candidates for expert validation. \lm can also serve as a specification-time assistant, helping 3GPP contributors detect design flaws before they propagate into implementations. Future work will explore cross-specification and cross-release analysis, broader threat models, improved candidate prioritization, and automated test execution.

\section{Acknowledgments}
This work was supported in part by the National Science Foundation (NSF) under Grants CNS-2321416, CNS-2246050, and NAIRR-240456. Any opinions, findings, conclusions, or recommendations expressed in this material are those of the authors and do not necessarily reflect the views of the NSF. We also thank the shepherd and reviewers for their help in improving this material.

\clearpage
\bibliographystyle{ACM-Reference-Format}
\bibliography{sample-base}

\clearpage
\appendix

\clearpage

\section{New Vulnerabilities and Proof-of-concept Attacks}
\label{append:new_vulnerabilities_v1_v7}

\subsection{V3: Vulnerable DNS-Based PLMN Discovery over Non-3GPP Access}

\noindent 3GPP TS~23.501 defines the support for non-3GPP access, including the non-roaming architecture, Local Breakout (LBO) architecture, and Home-Routed (HR) roaming architecture. In these architectures, a UE can attach via an untrusted non-3GPP access network (e.g., untrusted satellite, Wi-Fi) in a visited network. To integrate such access, the UE's traffic is routed through the Non-3GPP Interworking Function (N3IWF) associated with the appropriate Public Land Mobile Networks (PLMN), enforcing access control and mobility anchoring. 

To bootstrap 5G connectivity over untrusted non-3GPP access, the UE must discover an appropriate N3IWF. TS~23.501 specifies a DNS-based PLMN Discovery for non-3GPP access, where the UE performs a DNS NAPTR query to obtain service instance identifiers in the form of `<MNC>.<MCC>.pub.3gppnetwork.org'; the UE then uses DNS to resolve the IP addresses of N3IWF. This design is attractive from a deployment perspective because it reuses widely available DNS infrastructure without requiring pre-provisioned, static endpoint lists hardcoded on the UE. 
However, it also introduces a DNS attack surface. Prior measurements report that 88\% of organizations have experienced DNS attacks~\cite{PaloAltoDNSAttack}.
Importantly, the specification does not mandate authenticated DNS transport or DNSSEC validation. Consequently, the DNS-based PLMN discovery responses can be spoofed or manipulated and an adversary can steer the victim toward the attacker-controlled or unreachable endpoints. 
While subsequent N3IWF procedures include cryptographic protection and authentication, this discovery-stage misdirection can create meaningful denial-of-service by preventing the UE from connecting to a legitimate N3IWF. 

\begin{figure}[H]
    \centering
    \includegraphics[page=3,width=0.8\linewidth]{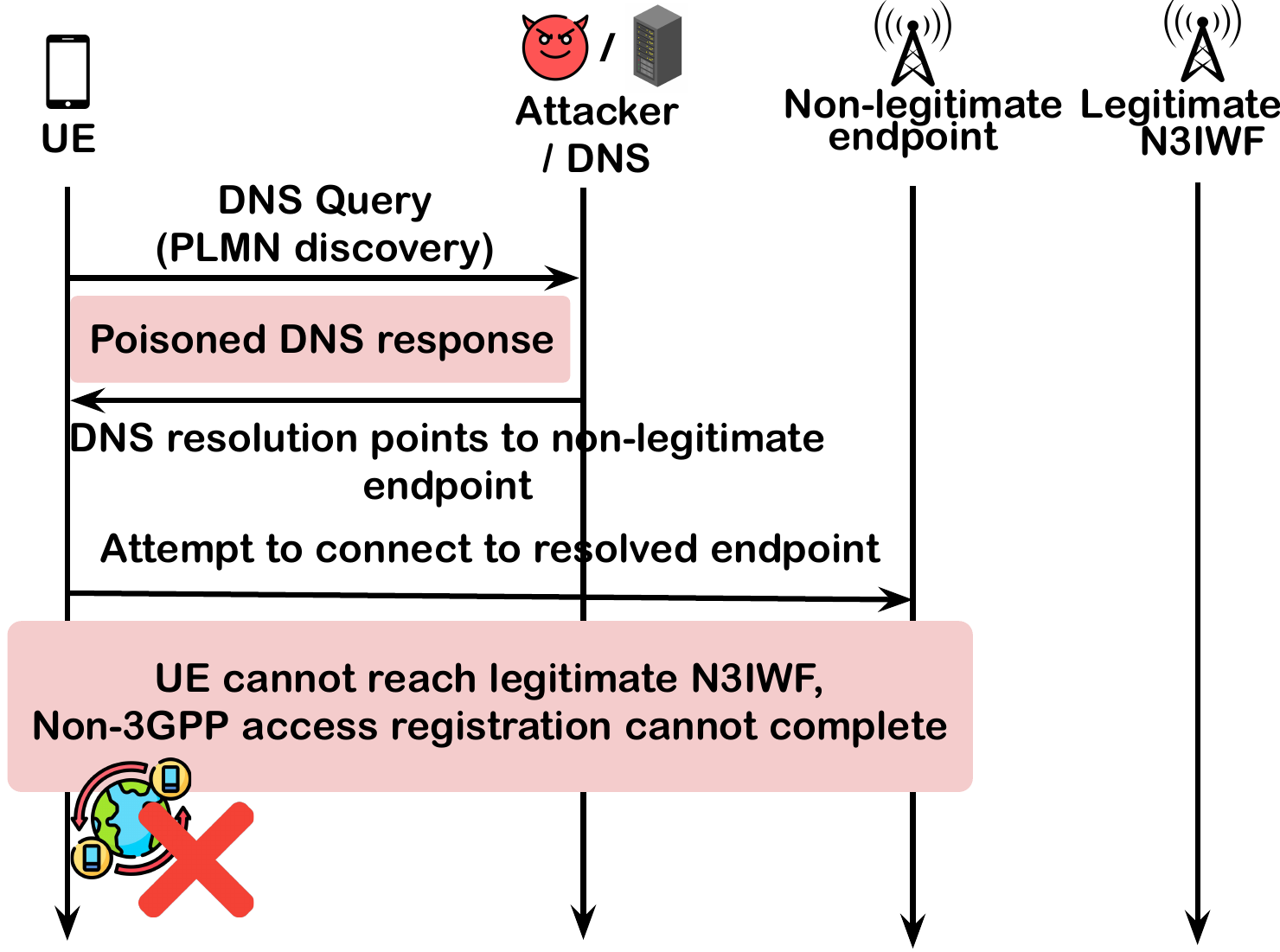}
    \vspace{-0.15in}
    \caption{Vulnerable DNS-Based PLMN Discovery over Non-3GPP Access}
    \label{fig:dns-base}
    \vspace{-0.2in}
\end{figure}

\subsubsection{Proof-of-concept Attack.} We validate this vulnerability from two perspectives: (1) whether a DNS security mechanism (i.e., DNSSEC) is deployed on the operational domain name to protect the DNS-based PLMN Discovery mechanism, and (2) whether a DNS manipulation attack can practically prevent a UE from connecting to the legitimate N3IWF, as illustrated in Figure \ref{fig:dns-base}. 

\begin{figure}[t]
    \centering
    \includegraphics[width=0.9\linewidth]{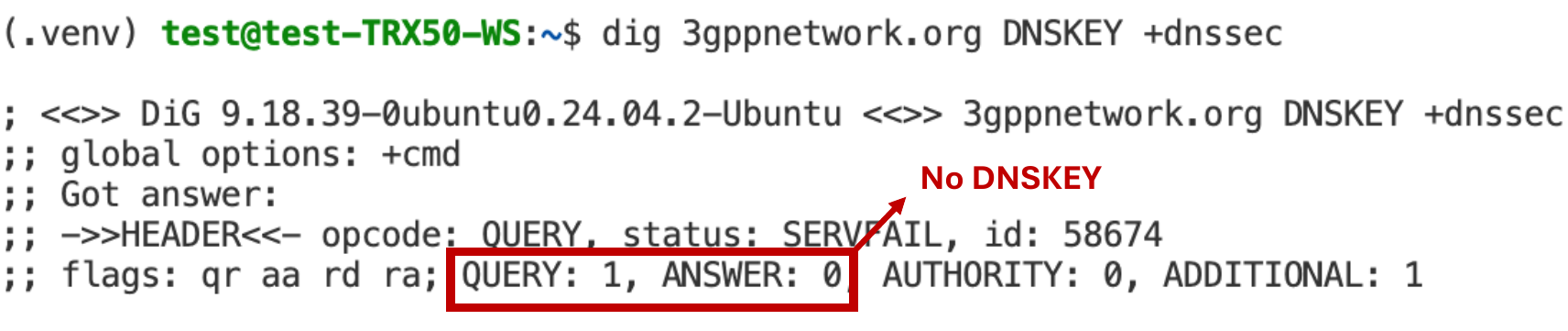}
    \vspace{-0.15in}
    \caption{`3gppnetwork.org' does not enable DNSSEC}
    \label{fig:dig-command}
    \vspace{-0.2in}
\end{figure}

For (1), we first test whether the parent domain `3gppnetwork.org', managed by GSMA, is protected by DNSSEC, because DNSSEC validation for any descendant name (e.g., DNS-based PLMN Discovery) relies on a valid chain of trust. 
 
Our measurement via the tool \texttt{dig}~\cite{isc-bind}, as shown in Figure~\ref{fig:dig-command}, indicates that the DNSSEC for `3gppnetwork.org' is absent. 
Therefore, DNS-based PLMN discovery remains vulnerable to classic DNS manipulation threats on untrusted links, even when subsequent procedures are cryptographically protected.  

For (2), to evaluate whether the DNS manipulation can work, we test whether a poisoned discovery response can prevent the UE from establishing a secure tunnel to the legitimate N3IWF. 
Since conducting Internet-scale poisoning experiments is unethical, we perform experiments in a controlled environment. 
Specifically, we use two COTS UEs, Samsung S24 and Moto G Stylus 5G on T-Mobile network, and connect them to a controlled Wi-Fi router configured to poison DNS queries relevant to `3gppnetwork.org'. After enabling Wi-Fi calling services, we observe that both UEs try to resolve the IP address of N3IWF (i.e., ePDG for Wi-Fi calling services). 
Under the poisoned resolution, the UEs attempt to connect to the misdirected endpoint, thereby preventing the UE from reaching the legitimate N3IWF and completing non-3GPP access registration. This shows that, even though subsequent N3IWF procedures provide cryptographic protection and mutual authentication, compromising the discovery stage can still induce a denial-of-service attack.  

\subsubsection{Root Cause and Lesson Learned.} The root cause is a trust gap in the DNS-based PLMN discovery, which is performed over untrusted access links before the UE establishes a secure channel to the operator's infrastructure. 
It is a standard design flaw that 3GPP specifications only protect the post-discovery connection and do not enforce security mechanisms during the discovery stage, which allows adversaries to exploit unauthenticated discovery to misdirect the victim.  
To address this vulnerability, future 3GPP specifications must enforce the security mechanisms on DNS-based PLMN discovery in two ways. First, discovery infrastructure must be treated as security-critical. Operational domains used for standardized discovery must provide an authenticated naming and resolution path. Second, standards need to mandate a robust fallback mechanism that avoids futile attempts.  

\begin{figure}[t]
    \centering
    \includegraphics[page=4,width=0.8\linewidth]{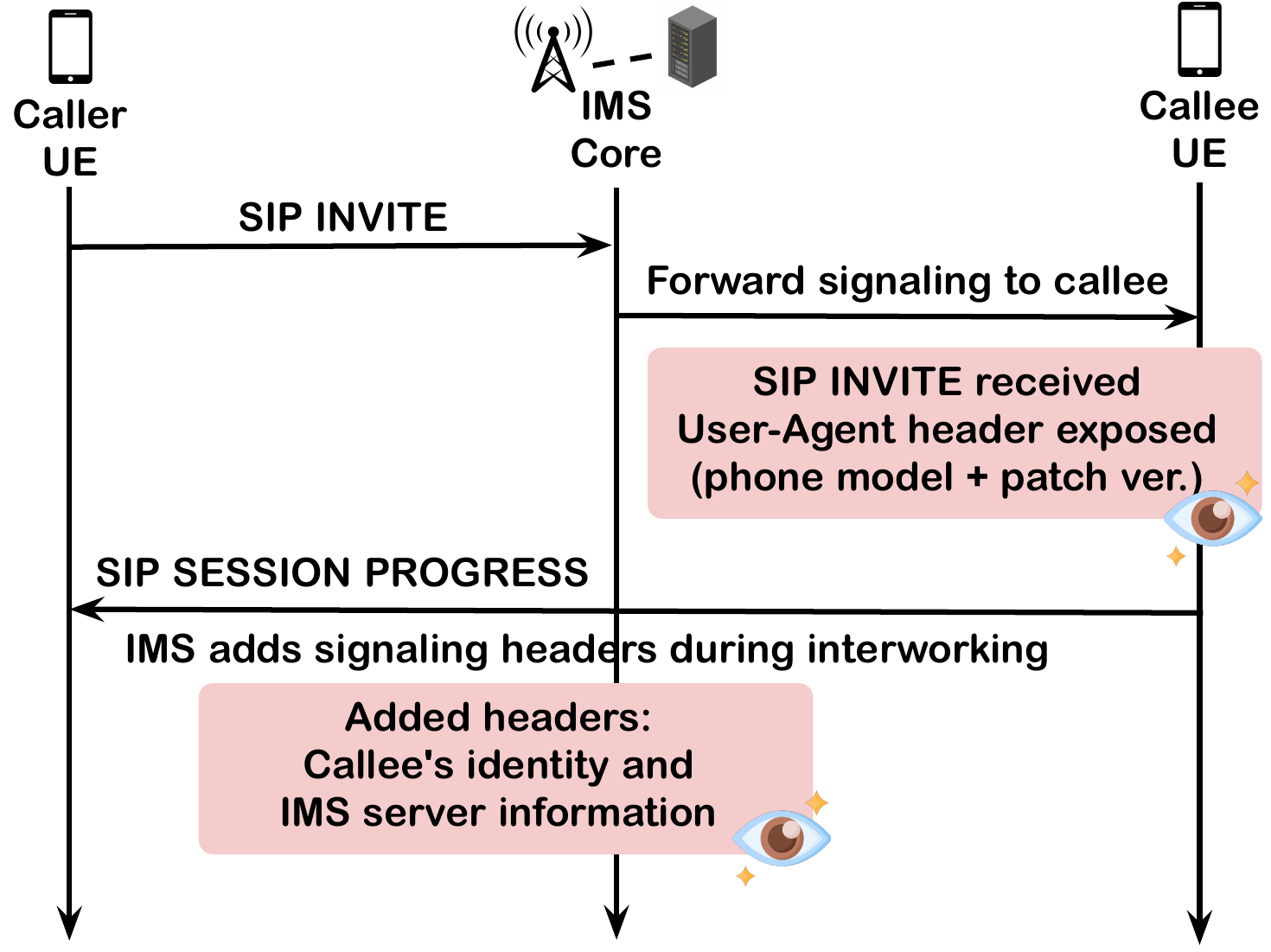}
    \vspace{-0.15in}
    \caption{Privacy Exposure via Over-Disclosed VoIMS Signaling Headers}
    \label{fig:privacy_exposure}
    \vspace{-0.2in}
\end{figure}

\subsection{V4: Privacy Exposure via Over-Disclosed VoIMS Signaling Headers} 

\noindent With the introduction of the all-IP network architecture in 4G LTE, call services previously supported by conventional circuit-switched (CS) technology in legacy 2G/3G networks have transitioned to being provided by the IMS. To facilitate IMS-based call services (also known as VoIMS - Voice over IMS), the Session Initiation Protocol (SIP), widely used in voice over IP services, was adopted for call signaling, replacing the legacy CS Call Control (CC) protocol. SIP is a text-based, string-oriented protocol, in contrast to the binary, structure-oriented CS CC protocol. 

Adopting a text-based call signaling protocol is a double-edged sword. While it offers greater flexibility and extensibility to support various emerging IP-based multimedia services compared to conventional protocols, it also introduces new issues due to potential missteps in design. Specifically, all 2G/3G CC signaling messages are designed to carry specific 3GPP-stipulated content only. In contrast, the text-based SIP protocol allows carriers and mobile devices to transmit a much broader range of content. 
Specifically, TS~24.229 stipulates 113 SIP headers and 15 SDP headers to support IMS-based call services, far more than its predecessor. These IMS service signaling headers can carry personal, device, and infrastructure details. 

Unfortunately, as shown in Figure \ref{fig:privacy_exposure}, there are no privacy preservation mechanisms enforced in 3GPP specifications for IMS signaling headers. While the signaling messages for normal IMS services are encrypted by IPSec, operators and UEs would carry such privacy-exposing headers to both IMS service users who can be the adversaries who collect the privacy information that may enable subsequent attacks.

\subsubsection{Experimental Validation.} We validate this potential privacy leakage by examining privacy leaks during call interworking. We control 2 phones, a Samsung S21 and a Google Pixel 5, on three US carriers (labeled as OP-I, OP-II, and OP-III) and collected signaling messages exchanged between two call ends. We make 10 calls for each combination of phone model and carrier. Our experimental results show that on the operational IMS services, privacy-leaking headers are transmitted and received by the UEs. Two privacy issues are explicitly identified in the messages SIP INVITE and SIP SESSION PROGRESS. In the former, as shown in Figure~\ref{fig:user-agent}, it is observed that the caller's User-Agent header is transmitted to the callee through the OP-II network, leaking the phone model and security patch version to the callee. For the latter as shown in Figure~\ref{fig:name-server}, when the caller is on OP-III, and the callee is on OP-I, P-Asserted-Identity and Server headers are added by the networks in the SIP SESSION PROGRESS response, revealing the callee's identity and IMS server information. 

\begin{figure}[t]
    \centering
    \begin{subfigure}[b]{0.45\textwidth}
        \centering
        \includegraphics[width=\textwidth]{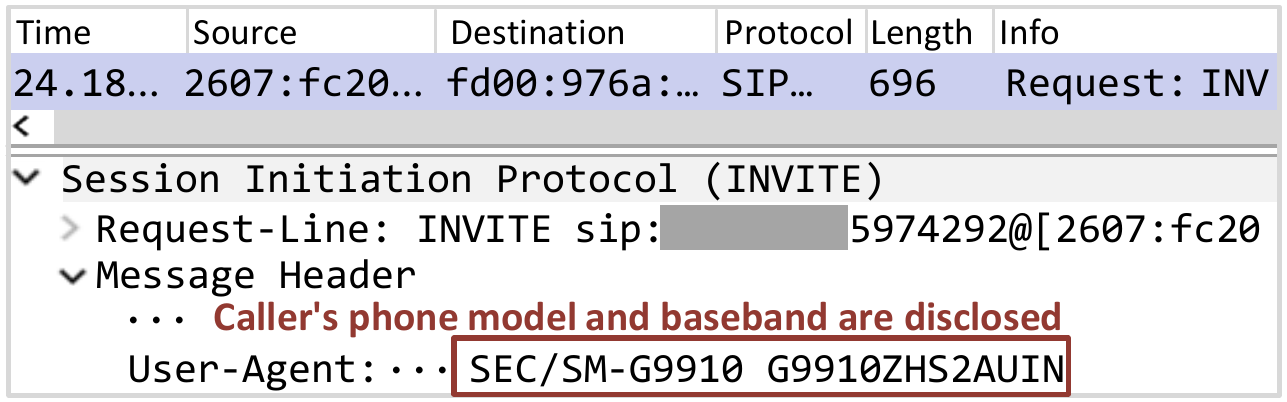}
        \caption{User-Agent header leaks phone model and security patch version}
        \label{fig:user-agent}
    \end{subfigure}
    \hfill
    \begin{subfigure}[b]{0.45\textwidth}
        \centering
        \includegraphics[width=\textwidth]{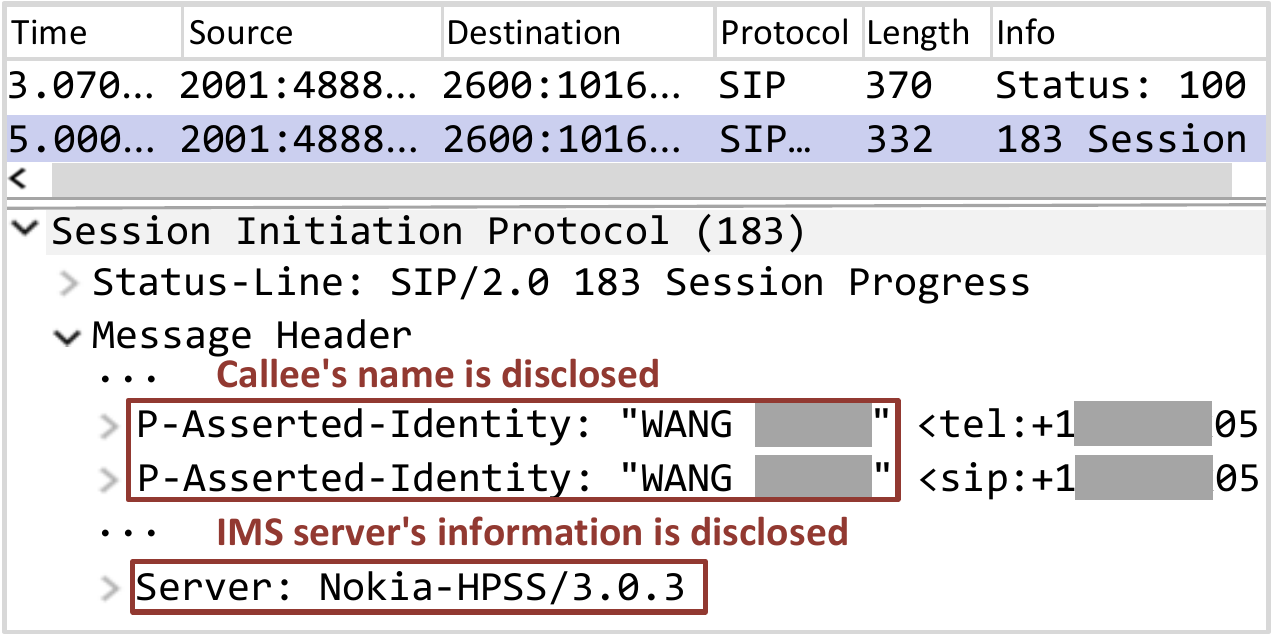}
        \caption{P-Asserted-Identity and Server headers leak callee's name and IMS server's information}
        \label{fig:name-server}
    \end{subfigure}
    \vspace{-0.1in}
    \caption{The information about users and cellular infrastructure is leaked during call interworking}
    \label{fig:main}
    \vspace{-15pt}
\end{figure}

\subsubsection{Root Cause and Lesson Learned.} Adopting a text-based call signaling protocol allows carriers and mobile devices to transmit a much broader range of content. While the 3GPP specification does not prohibit the transmission of these abundant headers, other network standard organizations like the IETF have raised concerns about privacy risks from unnecessary header disclosures. For example, RFC 3261~\cite{RFC3261} warns about the vulnerability of disclosing user agent information. This underscores the urgent need for 3GPP to review and revise its IMS-related specifications with a focus on privacy preservation.

\subsection{V5: Exploitable CS-Fallback Initiation} 

\begin{figure}[t]
    \centering
    \includegraphics[page=5,width=0.8\linewidth]{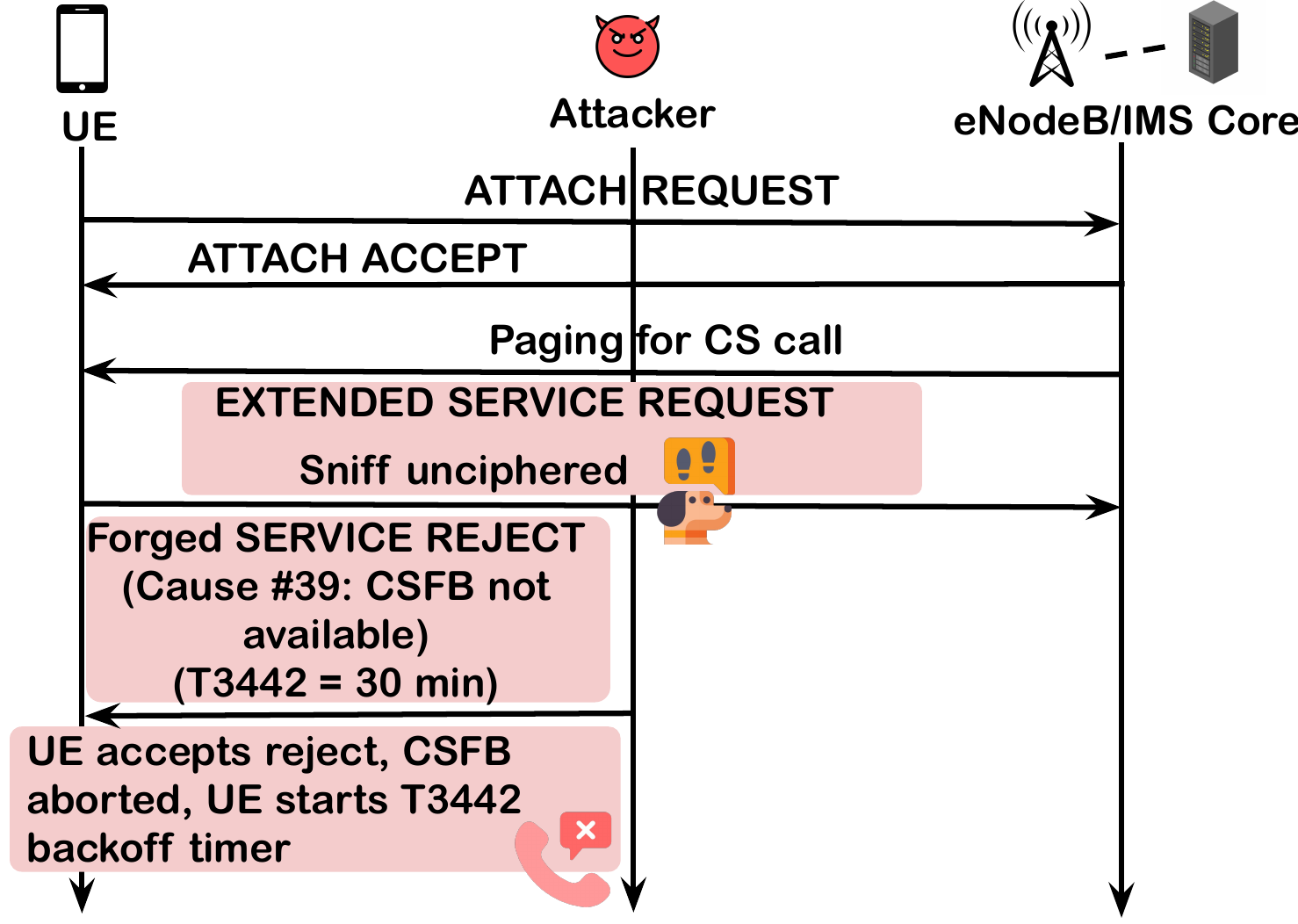}
    \vspace{-0.15in}
    \caption{Exploitable CS-Fallback Initiation}
    \label{fig:csfallback}
    \vspace{-0.2in}
\end{figure}

\noindent 4G LTE networks support Circuit-Switched Fallback (CSFB) to provide legacy CS voice/SMS by redirecting a UE to a CS network (e.g., 2G/3G if available) when VoLTE is unavailable. 
Note that 5G non-standalone deployments that rely on 4G LTE infrastructure also support CSFB.   
In CSFB, after a UE in EMM-IDLE receives a paging request for a CS service, it initiates fallback by sending an EXTENDED SERVICE REQUEST to the Core Network. In the common case, the UE transitions from RRC\_IDLE to RRC\_CONNECTED and sends this request as an initial NAS message. In this case, the initial message is only integrity-protected without ciphering because NAS ciphering is only applied after the access-stratum connection and the relevant security context are re-activated for this specific exchange. Thus, adversaries can sniff the \textsf{EXTENDED SERVICE REQUEST}. 

Moreover, TS~24.301 allows certain NAS reject messages, including SERVICE REJECT, to be sent and processed without any security protection. This combination exposes a practical CSFB DoS attack surface: adversaries can deny CSFB and rate-limit subsequent attempts by (1) sniffing unciphered EXTENDED SERVICE REQUEST and (2) injecting forged SERVICE REJECT with EMM cause \#39 (CS domain temporarily not available) and the associated timer T3442 to throttle subsequent attempts.

\subsubsection{Proof-of-concept Attack.} Figure~\ref{fig:csfallback} illustrates a proof-of-concept attack. The attacker deploys a sniffer between the victim and the legitimate eNodeB. Once an unciphered EXTENDED SERVICE REQUEST is observed, the attacker immediately sends a forged SERVICE REJECT carrying EMM cause \#39 and the T3442 via a rogue base station.  
To validate this attack, as shown in Figure~\ref{fig:device-setup}, we implement an SDR-based testbed using srsRAN and open5GS~\cite{open5gs}, both of which support CS Fallback. 
A Samsung S24 and a Moto G Stylus are used as test UEs. Instead of building a full injection pipeline that monitors unciphered NAS messages and injects forged responses, which are credited to prior works~\cite{hussain2018lteinspector}, we simplify the evaluation by configuring Open5GS and srsRAN to emulate a CS call paging and then directly respond with a SERVICE REJECT with EMM cause \#39 and T3442 for the maximum value 30 mins directly. 
As shown in Figure~\ref{fig:csfb}, the experimental traces indicate that the UEs accept the rejection and stop attempting to set up the call. 
Overall, this attack achieves a stealthy denial-of-service against CS services on LTE by preventing a CSFB procedure, and limiting further CSFB attempts via \textsf{T3442}. The throttling effect is particularly damaging because the UE self-enforces the backoff timer. Depending on UE implementation and timer configuration, the victim may experience prolonged inability to place CS calls. We leave a systematic characterization of the maximum outage duration to future work.

\begin{figure}[t]
    \centering
    \includegraphics[page=5,width=0.8\linewidth]{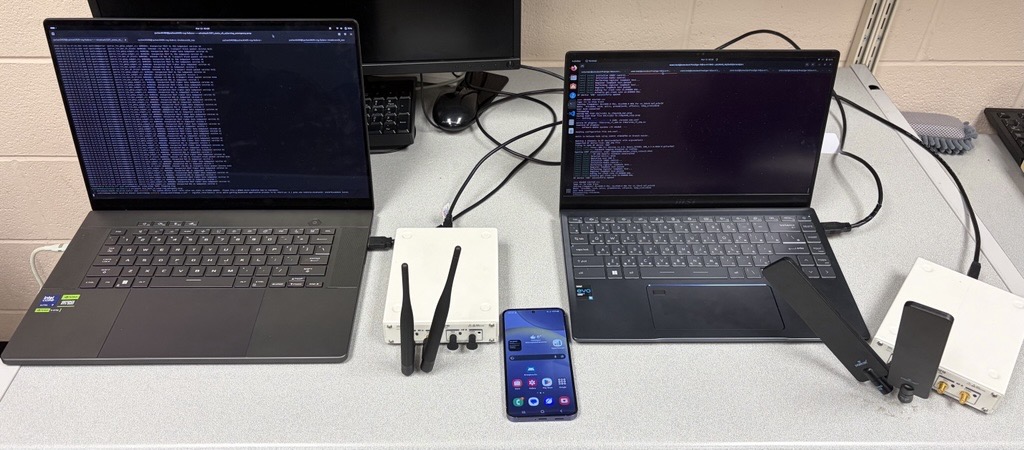}
    \vspace{-0.15in}
    \caption{The setup for validating V5: Exploitable CS-Fallback Initiation}
    \label{fig:device-setup}
    \vspace{-0.2in}
\end{figure}

\begin{figure}[t]
    \centering
    \includegraphics[page=5,width=0.95\linewidth]{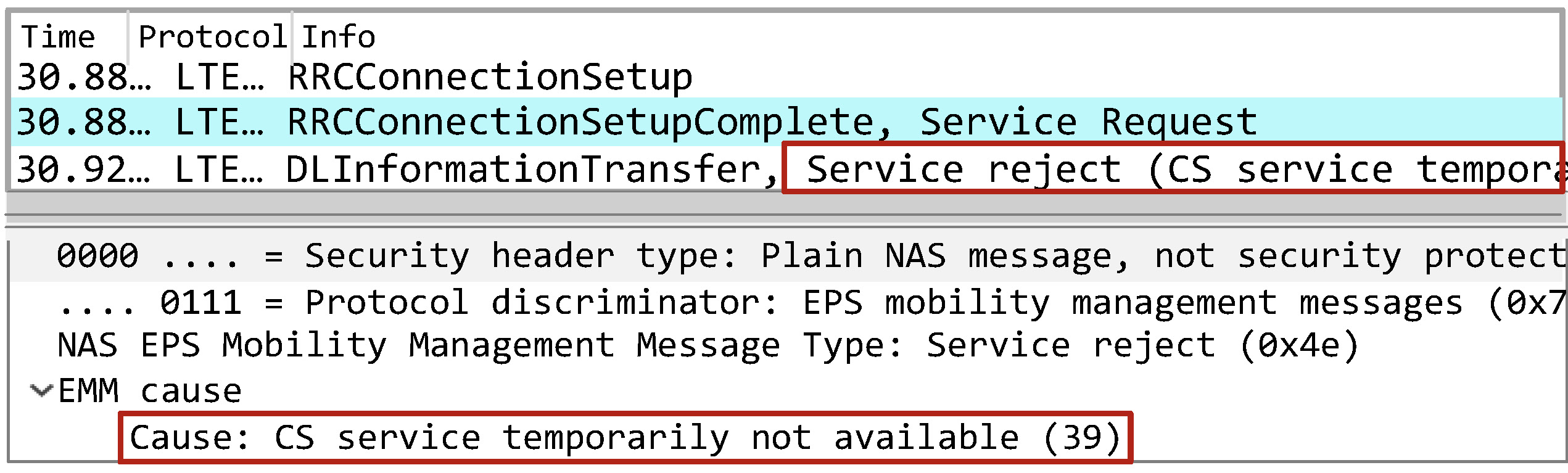}
    \vspace{-0.15in}
    \caption{The victim UE accepts the rejection and ceases its attempts to establish the call}
    \label{fig:csfb}
    \vspace{-0.2in}
\end{figure}

\subsubsection{Root Cause and Lesson Learned.} The root cause is TS~24.301 explicitly permits a SERVICE REJECT to be processed without ciphering and integrity protection. This design appears risky but serves as a fail-safe mechanism for cases where a secure connection cannot be established or has been lost. For example, if a UE sends an EXTENDED SERVICE REQUEST but the MME cannot authenticate the user (e.g., the unknown/invalid subscription), the network must be able to instruct the UE to stop trying without security keys. Similarly, for emergency services, the network must always be able to reject or redirect requests even if security procedures are failing to ensure the UE can seek alternative access. 
To mitigate this threat, commonly from rogue base stations, and still allow certain unprotected signaling messages for a fail-safe mechanism, in practice, it likely requires UE-side defenses against rogue base stations and stricter validation before setting the long throttling timers.

\subsection{V6: Access Trap via Access Barring Factor}

\noindent In 5G NR, System Information Block 1 (SIB1) is a periodically broadcast RRC message that every UE must decode first before any AS (Access Stratum) security is established. SIB1 carries essential cell-access and cell-selection parameters, including PLMN, scheduling of other system information blocks, and uac-BarringInfo for UAC (User Access Control). uac-BarringInfo specifies barring rules for different access categories, distinguishing everyday services (e.g., regular voice/data and mobile-originated signaling) from special-case/high-priority categories (e.g., paging/mobile-terminated access and emergency access). A similar design also existed in 4G networks, which used the ac-BarringInfo field in SIB2 messages.

Each barring rule includes parameters such as barringFactor, which controls the probability that a UE is allowed to initiate an access attempt for a given category. Intuitively, a smaller barringFactor imposes a stronger restriction, while an extremely small value can cause most or nearly all UEs to defer their mobile-originated requests. In normal operation, base stations typically configure moderate barring parameters, or disable barring entirely, to preserve broad service availability and avoid unnecessarily suppressing legitimate access.

This design becomes exploitable when a malicious base station advertises an extremely restrictive barringFactor in SIB1/SIB2 in 5G/4G networks. After decoding the broadcast information, the UE enforces the barring policy locally. By setting barringFactor to a highly restrictive value, the attacker can effectively drive victim UEs into a stealthy nonfunctional state, where the device appears camped on a cell but repeatedly refrains from initiating normal services, thereby creating a stealthy denial-of-service condition.

\begin{figure}[t]
    \centering
    \includegraphics[page=6,width=0.8\linewidth]{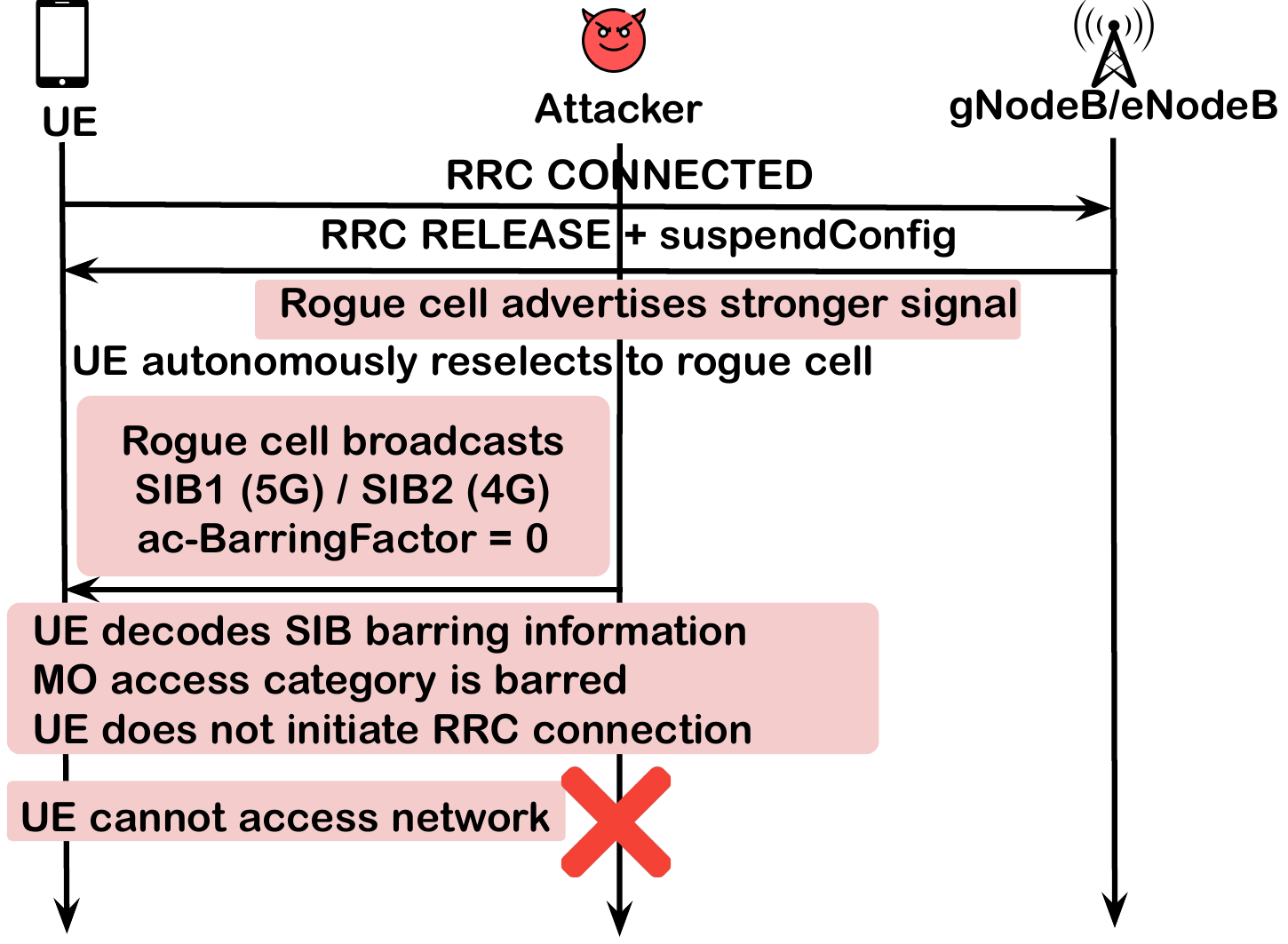}
    \vspace{-0.15in}
    \caption{Access Category Trap via SIB1 Access Barring}
    \label{fig:sib1}
    \vspace{-0.2in}
\end{figure}

\subsubsection{Proof-of-concept Attack.} Figure~\ref{fig:sib1} illustrates the proof-of-concept attack. The adversary only needs to operate a rogue base station and exploit broadcast system information; no cryptography is broken. Specifically, the adversary needs to configure the rogue base station to broadcast SIB1 in 5G/SIB2 in 4G with barringFactor set to a highly restrictive value (e.g., 0), which sets a low probability that a UE is allowed to initiate an access attempt. Therefore, any nearby UEs that receive and decode this SIB1/SIB2 are not able to initiate an access attempt.

To validate this vulnerability ethically, we implement an SDR-based testing platform using srsRAN and srsEPC. All antennas are placed in a shield box to ensure that no unintended over-the-air transmissions occur.
We use a Samsung S24 and a Moto G Stylus 5G as test phones. 
Next, we simplify the experiment and configure the controlled serving cell to broadcast SIB2 with an ac-BarringFactor setting to 0. As shown in Figure~\ref{fig:sib2}, we observe (i) the UE does not initiate RRC connection establishment after receiving a SIB2 with BarringFactor setting to 0, and (ii) the data service on UEs is not usable without interfaces. Our experimental results show that both phones self-block service without explicit alerting beyond generic `no service' icon, consistent with the intended UAC enforcement behavior but enabling a stealthy denial-of-service.

\begin{figure}[t]
    \centering
    \includegraphics[width=0.95\linewidth]{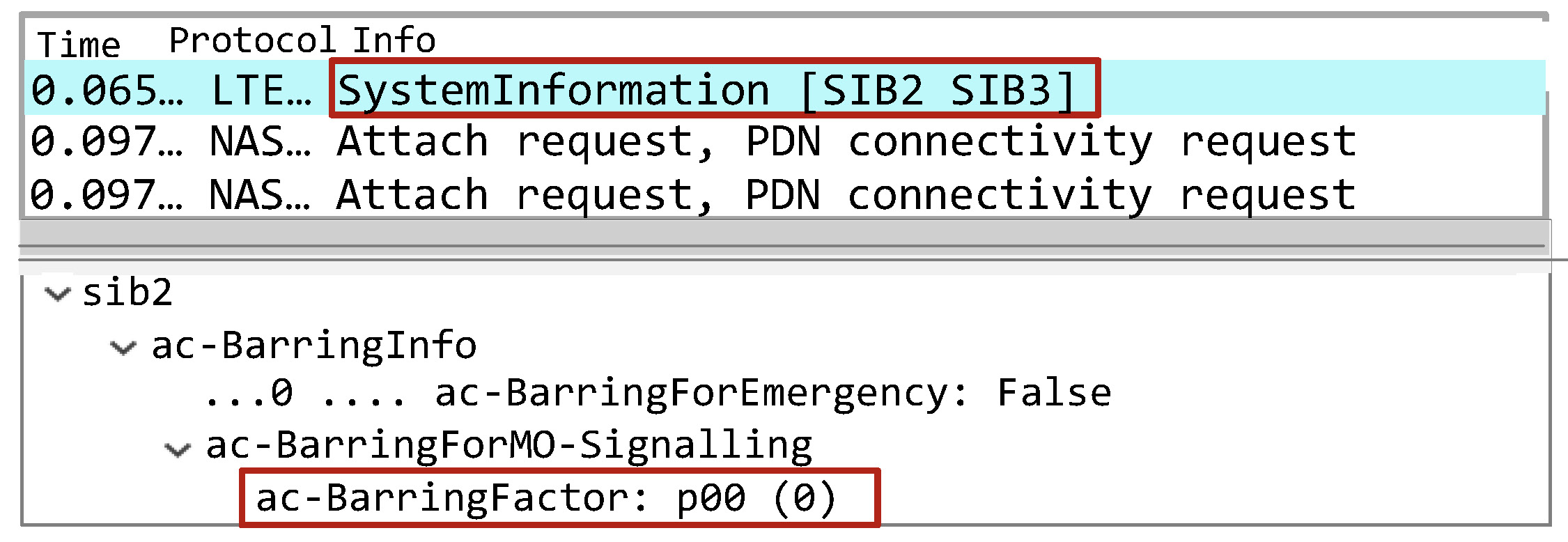}
    \vspace{-0.15in}
    \caption{The SIB2 messages broadcast by attackers prevent UEs from attaching to base stations}
    \label{fig:sib2}
    \vspace{-0.2in}
\end{figure}

\subsubsection{Root Causes and Lesson Learned.} UAC is designed for congestion control: the UE locally enforces access restrictions to reduce network load and stabilize access attempts during overload. However, the policy is delivered via SIB1/SIB2, which is broadcast and processed prior to authentication and therefore lacks integrity protection. This creates a pre-authentication trust gap: a rogue cell can advertise restrictive UAC policies and rely on the UE to self-enforce them, achieving denial-of-service without needing to defeat cryptographic protections.
To mitigate this vulnerability, 3GPP standards shall develop UE implementations that can reduce exposure by adding safety checks and recovery logic (e.g., bounding the duration of extreme barring, triggering reselection/blacklisting when a newly camped cell persistently bars most categories, or cross-validating with neighboring cells).

\subsection{V7: Rogue Base Station RRC RESUME Blackhole}

\noindent In 5G NR, a UE selects the cell based on the broadcast radio information and measured radio conditions before any AS/NAS security context is activated. Concretely, parameters conveyed via the broadcast messages (e.g., SSB/PBCH, MIB, and SIBs) are not authenticated at the UE during this procedure. As a result, a rogue base station can attract a UE through (1) stronger radio conditions in terms of RSRP, RSRQ, and SINR, (2) advertising the legitimate PLMN info that the UE is willing to use, and (3) advertising that it is not barred for achieving a high priority in the standard cell selection and reselection mechanism defined in TS~36.304.

Combining the cell selection and reselection mechanism with the RRC RESUME procedure that a UE in RRC INACTIVE expects to resume quickly via RRC RESUME REQUEST, the UE can be induced to reselect into a gNB that does not recognize or even a rogue gNB that can intentionally ignore or reject this request. 
This resume thus fails. Moreover, instead of directly switching to another cell, TS~36.304~\cite{3gpp_ts36304} defines that a UE should retry after a wait time specified by the timer T302. The cell selection and reselection mechanism does not consider the resume failure. Thus, the victim UE will stick to the malicious base station until a legitimate gNB wins in the cell selection and reselection procedure. 
While 5G security prevents a rogue gNB from completing an authenticated service, repeated resume failures can still cause meaningful availability harm for the retry loops.

\begin{figure}[t]
    \centering
    \includegraphics[page=7,width=0.8\linewidth]{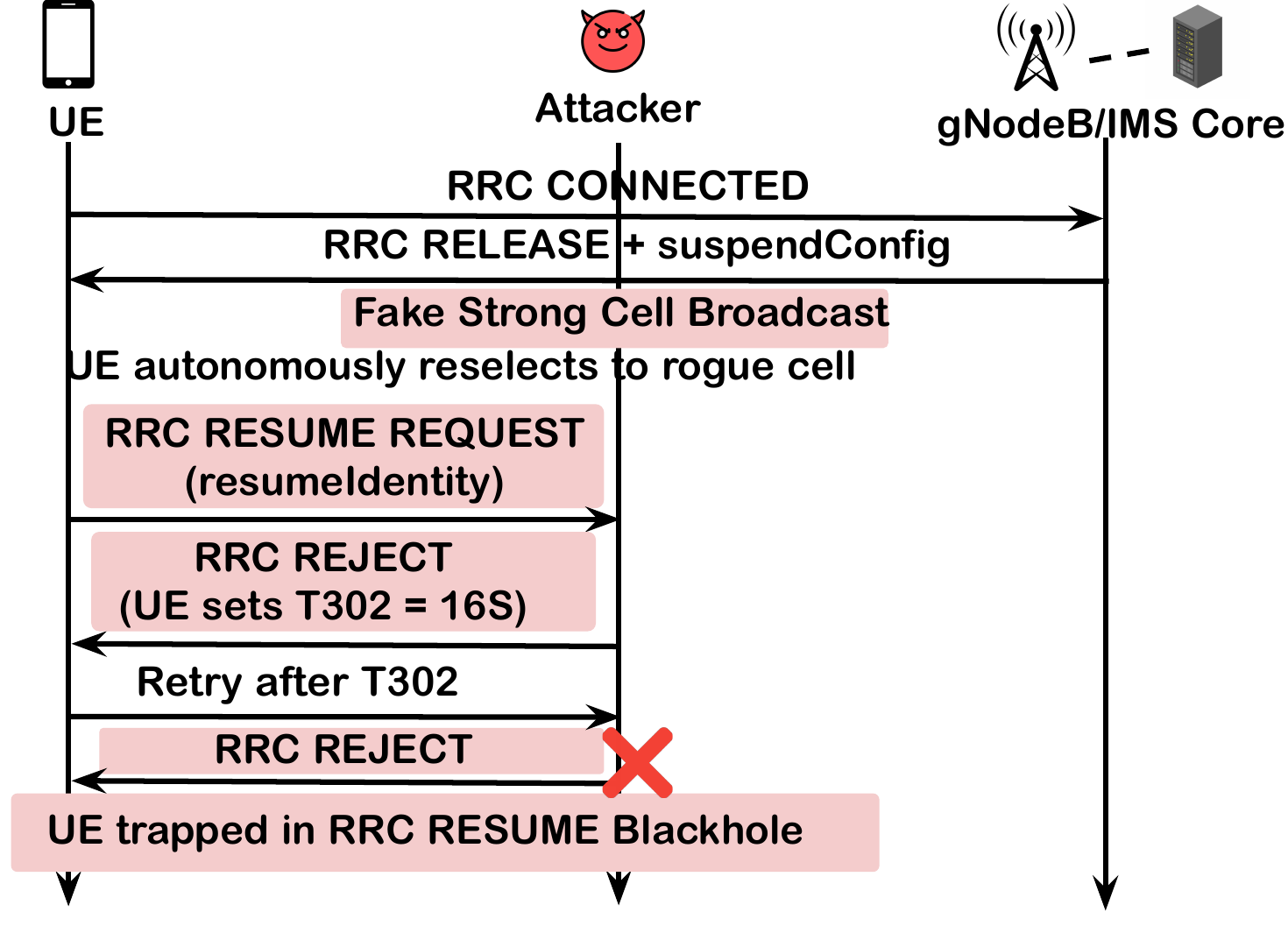}
    \vspace{-0.15in}
    \caption{Rogue Base Station RRC RESUME Blackhole}
    \label{fig:rrcresume_blackhole}
    \vspace{-0.2in}
\end{figure}

\begin{table*}[!t]
\centering
\caption{Comparison of existing systems for 3GPP specification security analysis}
\label{tab:spec-compare}
\vspace{-0.1in}
\scriptsize
\begin{tabular}{p{1.8cm} p{3cm} >{\centering}p{1.8cm} p{2.8cm} p{6cm}}
\toprule
\textbf{Method} & \textbf{Primary Objective} & \textbf{Protocol Modeling} & \textbf{Security Scope} & \textbf{Remarks} \\
\midrule

\textbf{Atomic}~\cite{chen2021bookworm} &
Detecting hazard indicators via textual entailment &
\xmark &
Risk‐sentence detection only &
Identifying hazard indicators via sentence-level textual entailment without modeling specification logic or performing systematic vulnerability analysis \\

\midrule

\textbf{ConTester}~\cite{chen2023sherlock} &
Extracting security requirements &
\xmark &
Developing conformance test cases &
Only extracting security requirements for developing conformance test without uncovering design-level vulnerabilities \\

\midrule

\textbf{CellularLint}~\cite{rahman2024cellularlint} &
Detecting semantic inconsistencies &
\xmark &
Sentence‐level contradiction detection &
Detecting semantic inconsistencies (i.e., sentence-level contradictions) via sentence-pair inference instead of reasoning over complete specification logic \\

\midrule

\textbf{Hermes}~\cite{al2024hermes} &
Automatic FSM extraction via traditional NLP approaches &
\cmark &
Formal security analysis &
Parsing specifications via handcrafted grammars using constituency parsing and dependency parse trees, remaining rule-intensive and lacking flexibility to adapt to evolving specifications \\

\midrule

\textbf{ARCANE}~\cite{tan2025automated} &
Automated vulnerability discovery via model-based fuzzing &
\cmark &
Implementation vulnerability discovery &
Automatically constructs protocol models and performs model-based fuzzing \\

\midrule

\textbf{\lm } &
End‐to‐end SCA nodes and systematic vulnerability reasoning &
\cmarkbold &
Specification‐level design‐flaw detection &
First system to enable scalable specification extraction and systematic security analysis \\

\bottomrule
\end{tabular}
\end{table*}

\subsubsection{Proof-of-concept Attack.} 
We refer to the following behavior as an \emph{RRC RESUME Blackhole}: the victim UE is trapped within a rogue gNB and stuck in repeated RRC RESUME REQUEST$\rightarrow$\ RRC RESUME failure cycles. 
The attacker does not need to break cryptography and does not need to complete mutual authentication. Instead, the attacker exploits (i) unauthenticated cell selection and (ii) the coupling between reselection and the RRC RESUME mechanism. 

\begin{figure}[H]
    \centering
    \includegraphics[width=0.95\linewidth]{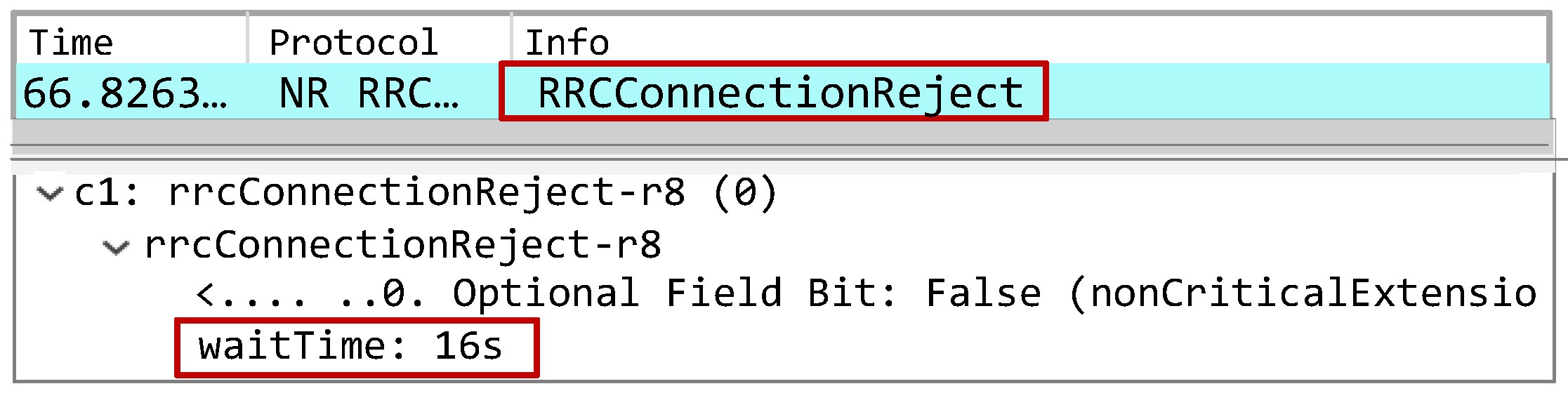}
    \vspace{-0.05in}
    \caption{The victim UE remains unable to resume services and repeats the procedure every 16 seconds according to the timer.}
    \label{fig:timer}
    \vspace{-0.2in}
\end{figure}

For ethical reasons, we validate this attack in a fully controlled environment. First, inside an RF shield box, we deploy two SDR-based gNBs using srsRAN and open5GS: One serves as a legitimate gNB, and another serves as the rogue gNB. The rogue gNB is configured to have same settings as the legitimate one. Two test phones, a Samsung S21 and a Moto G Stylus 5G, are employed. Second, we increase the rogue gNB signal strength via increasing its tx\_gain and rx\_gain in srsRAN so that the rogue gNB can be the first choice in the cell selection and reselection mechanism. Third, we operate the test phones to trigger the RRC RESUME. Fourth, we configure the rogue gNB to respond with RRC REJECT for all RRC RESUME REQUEST with a RejectWaitTime 16, which will be the time used by the UE as T302. Last, we examine if the UE is stuck to the rogue gNB and trapped within RRC RESUME Blackhole. Figure~\ref{fig:timer} illustrates a trace captured on the Moto G Stylus 5G, which remains unable to resume service for 16 seconds after receiving RRC REJECT. The Samsung S21 exhibits a different implementation behavior: it does not strictly follow the 3GPP-specified wait time and stops sending RRCResumeRequest for only about 3 seconds even when RejectWaitTime $=$ 16 is provided. This result shows that although the attack is effective on both devices, UE implementations may differ in how strictly they enforce the backoff timer.

\subsubsection{Root Causes and Lessons Learned.} The root cause is that the cell selection and reselection mechanism developed in 3GPP standards does not consider these failures. It operates on the RRC layer before the UE can establish a protected AS channel. Future specifications and deployments can consider additional mechanisms to reduce resume failures (e.g., explicit guidance for UEs to switch from cells that cannot support resumption).

\vspace{-5pt}

\section{Related Work}
\label{append:Related_Work}
Table \ref{tab:spec-compare} summarizes and contrasts representative 3GPP specification analysis systems in terms of their primary objectives, protocol modeling capabilities, security analysis scope, and key methodological characteristics.

\section{\textit{\lm}}
\label{append:prompt_algorithm}
This appendix provides implementation-level details of the prompts and algorithms used by \lm. It covers \textit{SpecAdaptation}, the \textit{SCA Representation Extractor}, the \textit{Function Chain Builder}, \textit{SecOracle}, and \textit{VulnTestGenerator}. The Function Chain Builder is formalized through Algorithm~\ref{alg:temporal_connection}, Algorithm~\ref{alg:semantic_connection}, Algorithm~\ref{alg:causal_connection}, and Algorithm~\ref{alg:reference_connection}, which respectively define Temporal, Semantic, Causal, and Reference Guided Connection.

\subsection{SpecAdaptation}

\vspace{-5pt}

\begin{algorithm}[ht]
\caption{\textit{SpecAdaptation}}
\label{alg:rag_prompt}
\scriptsize
\begin{algorithmic}[1]

\State \textbf{Input:} Versioned 3GPP corpus $\mathcal{D}$, component query $q$, prompt template $\mathcal{P}$, retrieval size $k$
\State \textbf{Output:} Grounded prompt $\tau$ for \core

\State \textit{/* Step 1: prepare a versioned specification index. */}
\State $\mathcal{D}' \leftarrow \textsc{Sanitize}(\mathcal{D})$
\State \textit{/* Remove formatting noise while preserving specification number, release/version, clause ID, and cross-references. */}

\State $\mathcal{I} \leftarrow \textsc{BuildIndex}(\mathcal{D}')$
\State \textit{/* Each indexed chunk keeps its source metadata for traceability. */}

\State \textit{/* Step 2: retrieve evidence for the current component. */}
\State $\mathcal{E} \leftarrow \textsc{RetrieveTopK}(q,\mathcal{I},k)$
\State \textit{/* The query may come from a clause, message type, a timer, an SCA node pair, security property, or candidate violation. */}

\State \textit{/* Step 3: expand evidence through explicit cross-references. */}
\State \textbf{for each} retrieved chunk $e \in \mathcal{E}$ \textbf{do}
    \State \quad $\mathcal{C}_e \leftarrow \textsc{ExtractReferencedClauses}(e)$
    \State \quad \textbf{if} $\mathcal{C}_e \neq \emptyset$ \textbf{then}
        \State \quad\quad $\mathcal{E} \leftarrow \mathcal{E} \cup \textsc{RetrieveClauses}(\mathcal{C}_e,\mathcal{I})$
        \State \quad\quad \textit{/* Include referenced clauses such as ``as specified in subclause X.Y.Z''. */}
    \State \quad \textbf{end if}
\State \textbf{end for}

\State \textit{/* Step 4: attach metadata for grounded reasoning. */}
\State $\mathcal{M} \leftarrow \textsc{GetMetadata}(\mathcal{E})$
\State \textit{/* Metadata includes specification number, release/version, clause ID, and source location. */}

\State \textit{/* Step 5: assemble the grounded prompt for the downstream module. */}
\State $\tau \leftarrow \textsc{AssemblePrompt}(\mathcal{P}, q, \mathcal{E}, \mathcal{M})$
\State \textit{/* The prompt contains task instructions, structured input, retrieved specification evidence, and required output schema. */}

\State \Return $\tau$

\end{algorithmic}
\end{algorithm}

\textit{SpecAdaptation} prepares the specification knowledge used by all later modules. It first sanitizes the collected 3GPP specifications by removing formatting noise while preserving clause numbers, cross-references, and release/version identifiers. It then builds a versioned retrieval index so that \core can reason over the relevant specification text rather than relying only on its parametric memory. Given a component query, such as a clause, an SCA node, a candidate node pair, or flagged violation, \core retrieves relevant specification excerpts from the indexed corpus and attaches them to the prompt together with clause identifiers and release/version information.

For example, when extracting an SCA node from the TS~24.501 sentence ``upon receipt of a message of a network-initiated 5GMM common procedure, the UE shall stop timer T3540 and respond to the network-initiated 5GMM common procedure via the existing N1 NAS signalling connection as specified in subclause 5.4,'' \textit{SpecAdaptation} retrieves the source clause containing this sentence and the referenced subclause 5.4. The retrieved evidence allows \core to ground the extraction in the analyzed TS~24.501 version, infer that the UE has an active N1 NAS signalling connection and timer T3540 is running, and extract the action that the UE stops T3540 and responds via the existing N1 NAS signalling connection. Algorithm~\ref{alg:rag_prompt} summarizes this workflow.

\subsection{SCA Representation Extractor}

The \textit{SCA Representation Extractor} maps specification text into structured SCA nodes. Each node contains four fields: start state, condition, action, and end state. The prompt in Figure~\ref{fig:sca_extractor_prompt} instructs \core to extract these fields from the retrieved specification evidence. If the specification text does not explicitly state a field, \core is instructed to infer it only when the surrounding context supports the inference; otherwise, the field is marked as not explicitly defined. This design preserves both explicit specification semantics and traceable best-effort inference.

\begin{figure}[ht]
    \centering
    \includegraphics[width=0.99\linewidth, page=1]{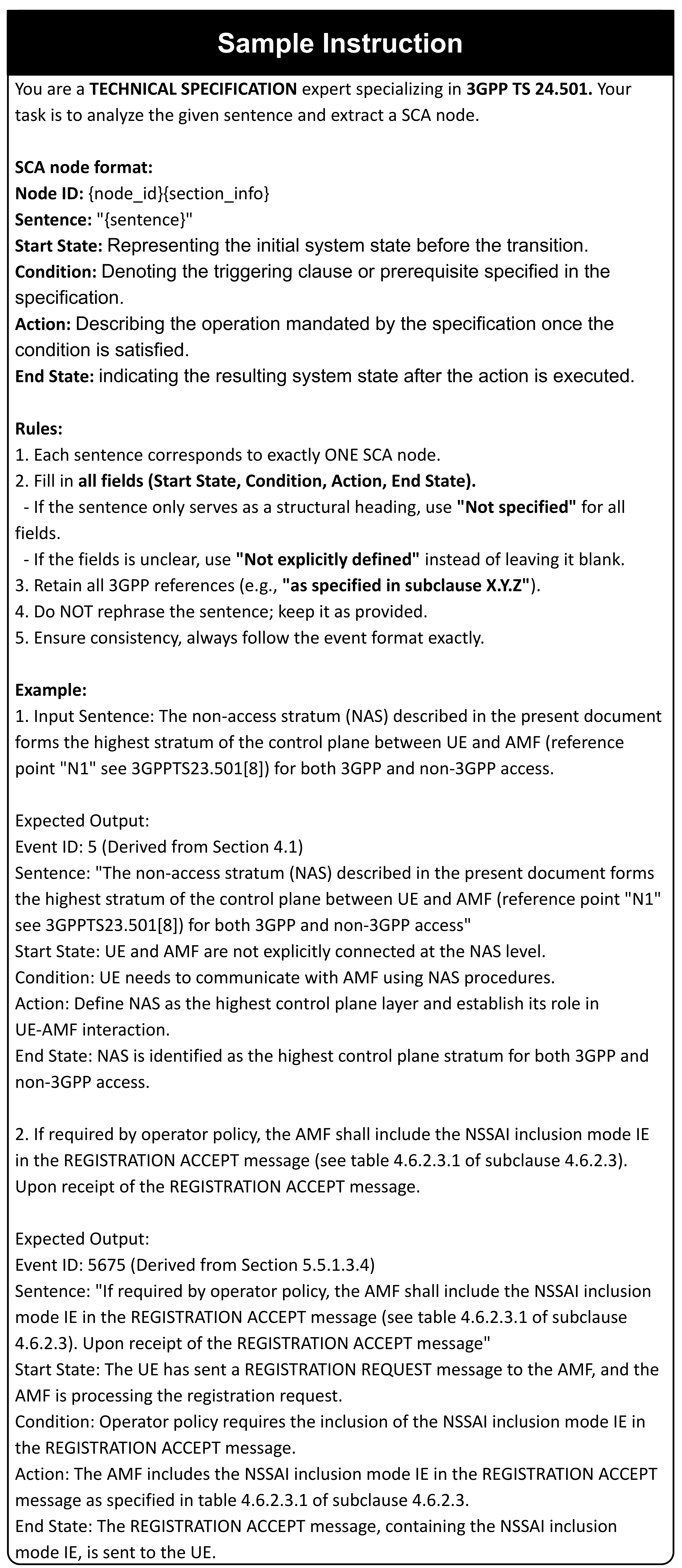}
    \vspace{-0.1in}
    \caption{ICL instruction for \textit{SCA Representation Extractor}}
    \vspace{-0.2in}
    \label{fig:sca_extractor_prompt}
\end{figure}

\subsection{Function Chain Builder}

After SCA nodes are extracted, the \textit{Function Chain Builder} connects scattered nodes into function chains. Temporal Connection links two nodes when the end state of one node exactly matches the start state of another node, as formalized in Algorithm~\ref{alg:temporal_connection}. Semantic Connection links two nodes when their end/start states are textually different but semantically equivalent, as described in Algorithm~\ref{alg:semantic_connection}. Causal Connection links two nodes when one node's state, condition, or action enables another node, as shown in Algorithm~\ref{alg:causal_connection}. Reference Guided Connection uses explicit cross-references in 3GPP specifications to narrow the search space before applying the above checks, as detailed in Algorithm~\ref{alg:reference_connection}.

\subsection{SecOracle}

Given a function chain and an adversarial action, the prompt in
Figure~\ref{fig:secoracle} asks \textit{SecOracle} to determine whether dropping, modifying, injecting, or replaying a message or event can produce a realistic candidate violation.

\noindent \textbf{Subject Assignment Verification.} We introduce a prompt-based actor and message-direction check before evaluating the security consequence. \textit{SecOracle} first identifies the attacked message and determines its sender and receiver from the corresponding SCA nodes and signaling direction. It then constrains receiver-side operations, such as processing, accepting, rejecting, validating, or handling a message, to the receiving entity rather than the sender.
Specifically, for a UE-to-network message, the consequence is analyzed at the receiving AMF, MME, or other network entity; for a
network-to-UE message, the consequence is analyzed at the receiving UE. If the message direction or involved entities cannot be determined from the specification evidence, \textit{SecOracle} is instructed to report the uncertainty rather than infer an unsupported actor.

The prompt also checks whether a valid security context, integrity
verification, freshness mechanism, and replay protection are explicitly active at the analyzed transition. This prevents \textit{SecOracle} from assuming that a message is protected merely because such protection exists elsewhere in the protocol.

The output is a structured candidate-violation report containing the
candidate-violation decision, potentially violated security
requirements, explanation, issue classification, validation test idea,
and validation source node.

\begin{algorithm}[ht]
\caption{Temporal Connection}
\label{alg:temporal_connection}
\scriptsize
\begin{algorithmic}[1]

\State \textbf{Input:} Set of SCA nodes $N=\{n_1,\dots,n_m\}$
\State \textbf{Output:} Temporal-Connection relation $T$

\State $T \leftarrow \emptyset$
\State \textit{/* Temporal connection is a strict state-matching rule. */}
\State \textit{/* It links $n_i$ to $n_j$ only when the end state of $n_i$ is the same as the start state of $n_j$. */}

\For{each ordered pair $(n_i,n_j) \in N \times N$}

    \State $e_i \leftarrow \text{end}(n_i)$
    \State $s_j \leftarrow \text{start}(n_j)$

    \State \textit{/* Skip underspecified states because they cannot support reliable temporal matching. */}
    \If{$e_i = s_j$ \textbf{and} $e_i \notin 
    \{\text{``not specified''},\text{``not explicitly defined''}\}$}

        \State $T \leftarrow T \cup \{(n_i,n_j)\}$
        \State \textit{/* $n_j$ is a valid temporal successor of $n_i$. */}

    \EndIf

\EndFor

\State \Return $T$

\end{algorithmic}
\end{algorithm}

\begin{algorithm}[ht]
\caption{Semantic Connection}
\label{alg:semantic_connection}
\scriptsize
\begin{algorithmic}[1]

\State \textbf{Input:} Set of SCA nodes $N=\{n_1,\dots,n_m\}$, model $\core$
\State \textbf{Output:} Semantic-Connection relation $S$

\State $S \leftarrow \emptyset$
\State \textit{/* Semantic connection handles cases where two states are equivalent but written differently. */}

\For{each ordered pair $(n_i,n_j) \in N \times N$}

    \State $e_i \leftarrow \text{end}(n_i)$
    \State $s_j \leftarrow \text{start}(n_j)$

    \If{$e_i$ and $s_j$ are valid}

        \State \textit{/* Step 1: Encode the two state descriptions. */}
        \State $\mathbf{t}_i \leftarrow \textsc{Tokenize}(e_i)$
        \State $\mathbf{t}_j \leftarrow \textsc{Tokenize}(s_j)$

        \State $\mathbf{H}_i \leftarrow \core.\textsc{Encode}(\mathbf{t}_i)$
        \State $\mathbf{H}_j \leftarrow \core.\textsc{Encode}(\mathbf{t}_j)$

        \State $\mathbf{z}_i \leftarrow \textsc{MeanPool}(\textsc{LastLayer}(\mathbf{H}_i))$
        \State $\mathbf{z}_j \leftarrow \textsc{MeanPool}(\textsc{LastLayer}(\mathbf{H}_j))$

        \State \textit{/* Step 2: Estimate textual semantic similarity. */}
        \State $sim \leftarrow \textsc{Cosine}(\mathbf{z}_i,\mathbf{z}_j)$

        \State \textit{/* Step 3: Ask \core to verify whether the two states represent the same protocol status. */}
        \State $\tau \leftarrow \textsc{AssemblePrompt}(\mathcal{P}_{sem}, e_i, s_j, sim)$
        \State $r \in \{\text{Yes},\text{No}\} \leftarrow \core(\tau)$

        \If{$r = \text{Yes}$}
            \State $S \leftarrow S \cup \{(n_i,n_j)\}$
            \State \textit{/* $n_j$ is linked as a semantic successor of $n_i$. */}
        \EndIf

    \EndIf

\EndFor

\State \Return $S$

\end{algorithmic}
\end{algorithm}

\begin{algorithm}[ht]
\caption{Causal Connection}
\label{alg:causal_connection}
\scriptsize
\begin{algorithmic}[1]

\State \textbf{Input:} Set of SCA nodes $N=\{n_1,\dots,n_m\}$, model $\core$
\State \textbf{Output:} Causal-Connection relation $C$

\State $C \leftarrow \emptyset$
\State \textit{/* Causal connection captures enabling relations that cannot be reduced to exact state matching. */}
\State \textit{/* For example, the action of one node may establish a context, timer, or connection required by another node. */}

\For{each ordered pair $(n_i,n_j) \in N \times N$}

    \State \textit{/* Step 1: Collect the full SCA fields of both nodes. */}
    \State $x_i \leftarrow \{\text{start}(n_i),\text{condition}(n_i),\text{action}(n_i),\text{end}(n_i)\}$
    \State $x_j \leftarrow \{\text{start}(n_j),\text{condition}(n_j),\text{action}(n_j),\text{end}(n_j)\}$

    \State \textit{/* Step 2: Build a reasoning prompt with the causal definition and decision rules. */}
    \State $\tau \leftarrow \textsc{AssemblePrompt}(\mathcal{P}_{causal}, x_i, x_j)$

    \State \textit{/* Step 3: \core decides whether $n_i$ enables, triggers, or makes $n_j$ reachable. */}
    \State $r \in \{\text{Yes},\text{No}\} \leftarrow \core(\tau)$

    \If{$r = \text{Yes}$}
        \State $C \leftarrow C \cup \{(n_i,n_j)\}$
        \State \textit{/* $n_j$ is linked as a causal successor of $n_i$. */}
    \EndIf

\EndFor

\State \Return $C$

\end{algorithmic}
\end{algorithm}

\begin{algorithm}[ht]
\caption{Reference Guided Connection}
\label{alg:reference_connection}
\scriptsize
\begin{algorithmic}[1]

\State \textbf{Input:} Set of SCA nodes $N=\{n_1,\dots,n_m\}$, clause-to-node index $\mathcal{M}$
\State \textbf{Output:} Reference Guided relation $R$

\State $R \leftarrow \emptyset$
\State \textit{/* Reference Guided Connection reduces the search space using explicit cross-references in specifications. */}
\State \textit{/* It avoids comparing all node pairs when the text already points to relevant clauses. */}

\For{each node $n_i \in N$}

    \State \textit{/* Step 1: Detect references such as ``as specified in subclause X.Y.Z''. */}
    \State $\mathcal{C} \leftarrow \textsc{ExtractReferencedClauses}(n_i)$

    \If{$\mathcal{C} = \emptyset$}
        \State \textbf{continue}
    \EndIf

    \State \textit{/* Step 2: Retrieve only nodes derived from the referenced clauses. */}
    \State $\mathcal{N}_c \leftarrow \textsc{RetrieveCandidateNodes}(\mathcal{C},\mathcal{M})$

    \State \textit{/* Step 3: Validate whether each referenced candidate forms a valid connection. */}
    \For{each candidate $n_j \in \mathcal{N}_c$}

        \If{\textsc{TemporalCheck}$(n_i,n_j)$ 
            \textbf{or}
            \textsc{SemanticCheck}$(n_i,n_j)$ 
            \textbf{or}
            \textsc{CausalCheck}$(n_i,n_j)$}

            \State $R \leftarrow R \cup \{(n_i,n_j)\}$
            \State \textit{/* A reference-guided edge is added only after passing at least one connection check. */}

        \EndIf

    \EndFor

\EndFor

\State \Return $R$

\end{algorithmic}
\end{algorithm}

\subsection{VulnTestGenerator}
After \textit{SecOracle} flags a candidate violation, \textit{VulnTestGenerator} converts the candidate-violation report into a validation-oriented test case. The prompt in Figure~\ref{fig:vulnTestGenerator} instructs \core to specify the required network state, UE state, security-context status, attacker capability, operation sequence, and expected observation. The generated test case does not by itself execute the experiment; instead, it provides a reproducible procedure that guides validation on a controlled testbed or operational network when ethical and practical constraints permit.

\begin{figure}[ht]
    \centering
    \includegraphics[width=1\linewidth, page=5]{pic/SCA_Representation_Extractor.pdf}
    \vspace{-0.1in}
    \caption{\textit{SecOracle}}
    \vspace{-0.1in}
    \label{fig:secoracle}
\end{figure}

\begin{figure}[ht]
    \centering
    \includegraphics[width=1\linewidth, page=6]{pic/SCA_Representation_Extractor.pdf}
    \caption{\textit{VulnTestGenerator}}
    \label{fig:vulnTestGenerator}
\end{figure}

\section{Implementation and Datasets} 
\label{append:implementation-datasets}
We use a workstation running Ubuntu 24.04 LTS with an NVIDIA GeForce RTX 4090 GPU (CUDA 12.6; NVIDIA driver 560.35.03) as the primary client environment to access remote compute resources and run supporting experiments. 4 LLMs, DeepSeek-V3.2~\cite{deepseek_v32_release_2025}, Qwen-Plus~\cite{qwen_plus_modelstudio_2026}, DeepSeek-R1-8B~\cite{deepseek_r1_distill_llama8b_2025}, and Qwen3-4B~\cite{qwen3_4b_instruct_2507} are used as the base models in the evaluation. We access the first two models through their public APIs. The later two are self-hosted on Jetstream2~\cite{jetstream2025llminference} using two H100s. DeepSeek-V3.2-Release is primarily used as the base model to build \core.  
The implementations for vulnerability validation differ across vulnerabilities; we describe the corresponding setups and procedures in the relevant vulnerability sections.

\noindent \textbf{Specification Dataset.} To ensure comparability with prior works, we construct a dataset from the same 3GPP specifications that govern control-plane signaling for NAS and RRC, which are critical for the security and mobility of operational 5G and 4G networks. In addition, to showcase that \lm can be applied to other 3GPP standards, specifications for 5G system architecture and SIP/SDP protocols are also included.
Thus, the dataset includes: \textbf{\textit{TS 38.331}} version 17.0.0 Release 17~\cite{3GPP_TS_38331}, \textbf{\textit{TS 24.501}} version 16.8.0 Release 16 \cite{3GPP_TS_24501}, \textbf{\textit{TS 24.301}} version 16.8.0 Release 16 \cite{3GPP_TS_24301}, \textbf{\textit{TS 23.501}} version 20.0.0 Release 20 \cite{3GPP_TS_23501}, and \textbf{\textit{TS 24.229}} version 19.5.0 Release 19 \cite{ETSI-TS-124229}.  

\clearpage

\section{Security Properties}
\label{append:securityproperties}

\subsection*{E.1 Authentication}
Ensures that only legitimate users, devices, and network entities can access services. \\
\noindent$\bullet$~\textbf{UE Authentication:} Ensures that only authorized users can access the network. \\ 
\noindent$\bullet$~\textbf{Network Entity Authentication:} Verifies the legitimacy of network nodes (eNodeB, gNodeB, AMF, SMF, etc.).  \\
\noindent$\bullet$~\textbf{Mutual Authentication:} Guarantees two-way authentication between the UE and the network to prevent man-in-the-middle attacks.  \\
\noindent$\bullet$~\textbf{SBA Authentication Binding:} Enforces token-based authentication between Service-Based Architecture (SBA) functions.  \\
\noindent$\bullet$~\textbf{Secure PLMN Selection:} Protects against forced redirection to rogue or unauthorized Public Land Mobile Networks (PLMNs).  \\
\noindent$\bullet$~\textbf{Roaming \& Inter-PLMN Security:} Ensures consistent authentication policies across operators during roaming.

\subsection*{E.2 Authorization}
Controls permissions and restricts operations to authorized entities only. \\
\noindent$\bullet$~\textbf{API \& Service Access Control:} Enforces strict authorization for APIs and service-based interfaces.  \\
\noindent$\bullet$~\textbf{Policy \& QoS Rule Protection:} Prevents unauthorized modification of policies, Quality of Service (QoS) rules, or charging rules.  \\
\noindent$\bullet$~\textbf{Edge Application Security:} Ensures strong isolation between Multi-access Edge Computing (MEC) applications.  \\
\noindent$\bullet$~\textbf{SBA API Security:} Requires strong authorization and input validation for Service-Based Architecture (SBA) REST APIs.  \\
\noindent$\bullet$~\textbf{Slice Isolation:} Ensures strict separation between 5G network slices (eMBB, URLLC, mMTC, etc.).  \\
\noindent$\bullet$~\textbf{Multi-Tenancy Isolation:} Guarantees tenant isolation in SBA and MEC environments.  \\
\noindent$\bullet$~\textbf{Network Function Isolation:} Enforces logical separation between core network functions (AMF, SMF, UPF, etc.).  \\

\subsection*{E.3 Service Integrity}
Guarantees that services, signaling, and data remain correct, unaltered, and trustworthy. \\
\noindent$\bullet$~\textbf{Message Integrity:} Ensures signaling and data messages are not altered during transmission.  \\
\noindent$\bullet$~\textbf{Security Context Validation:} Ensures NAS/RRC security contexts are properly established before any sensitive exchanges.  \\
\noindent$\bullet$~\textbf{Secure Paging Mechanisms:} Validates paging messages and prevents spoofed triggers.  \\
\noindent$\bullet$~\textbf{Secure Emergency Services:} Ensures integrity-protected emergency registration and deregistration procedures.

\subsection*{E.4 Service Confidentiality}
Protects sensitive signaling and user data from unauthorized access or disclosure.\\
\noindent$\bullet$~\textbf{Data Encryption:} Protects user data and signaling from interception. \\  
\noindent$\bullet$~\textbf{Service Confidentiality:} Ensures sensitive information remains protected at all times.  \\
\noindent$\bullet$~\textbf{Integrity \& Confidentiality Binding:} Ensures both encryption and integrity protections are activated together.  \\
\noindent$\bullet$~\textbf{Secure Algorithm Negotiation:} Prevents downgrades to weak ciphering or integrity algorithms.  \\
\noindent$\bullet$~\textbf{Session Key Update:} Secures key re-derivation and renewal during mobility or context switches.  \\
\noindent$\bullet$~\textbf{Key Lifecycle Management:} Ensures secure generation, distribution, update, and revocation of cryptographic keys.  

\begin{table*}[!t]
\scriptsize
\centering
\caption{Test case for validating \texttt{service integrity} violation in Node 2073}
\label{tab:replay-2073}
\vspace{-0.1in}
\setlength{\tabcolsep}{3pt}
\renewcommand{\arraystretch}{1.1}
\begin{tabular}{|c|p{7cm}|c|p{3.5cm}|p{3cm}|c|}
\hline
\multicolumn{6}{|p{0.96\textwidth}|}{
\textbf{Direction legend.} The U--M column indicates the signaling direction between the UE and the mobile network: $\rightarrow$ denotes a UE-to-network message, $\leftarrow$ denotes a network-to-UE message, and ``-'' denotes a local state transition or a step without an explicitly exchanged message.
} \\
\hline
\textbf{Step} & \textbf{Procedure} & \textbf{U--M} & \textbf{Message} & \textbf{Parameter} & \textbf{Verdict} \\
\hline
1 & The UE is switched on & - & - & - & - \\
\hline
2 & The UE initiates the registration procedure & $\rightarrow$ & REGISTRATION REQUEST & initial registration & - \\
\hline
3 & The network accepts the registration and UE enters 5GMM-CONNECTED & $\leftarrow$ & REGISTRATION ACCEPT & - & - \\
\hline
4 & The lower layer sends RRC suspend indication to UE & $\leftarrow$ & RRC Connection Suspend & - & - \\
\hline
5 & The UE transitions to 5GMM-CONNECTED with RRC inactive indication & - & - & - & - \\
\hline
6 & A trigger occurs to update UE radio capability; the UE prepares a new registration & $\rightarrow$ & REGISTRATION REQUEST & NG-RAN-RCU = ``UE capability update needed'' & - \\
\hline
7 & The UE transitions to 5GMM-IDLE and initiates the capability update registration & - & - & - & - \\
\hline
8 & The attacker captures the REGISTRATION REQUEST in transit & $\rightarrow$ & (Captured) REGISTRATION REQUEST & - & - \\
\hline
9 & The attacker replays the REGISTRATION REQUEST to the network & $\rightarrow$ & REGISTRATION REQUEST & - & - \\
\hline
10 & The AMF processes the replayed REGISTRATION REQUEST without sufficient authentication, causing the existing UE connection or registration context to be released, replaced, or disrupted & - & - & Existing UE context affected & Fail \\
\hline
\end{tabular}
\end{table*}

\subsection*{E.5 Privacy Protection}
Prevents leakage of sensitive identifiers, location information, and other personal data.\\
\noindent$\bullet$~\textbf{Identifier Protection:} Ensures SUPI, SUCI, GUTI, and other permanent identifiers are never exposed in plaintext.  \\
\noindent$\bullet$~\textbf{Location Privacy:} Protects against user tracking through paging, tracking area (TA) updates, or broadcast messages.  \\
\noindent$\bullet$~\textbf{Secure Paging Privacy:} Prevents identity leakage during paging in multi-network environments.  \\
\noindent$\bullet$~\textbf{Privacy Protection Mechanisms:} Ensures slice, subscription, and policy data remain confidential.  

\subsection*{E.6 Network Availability \& Signaling Security}
Ensures the continuous availability and reliability of network services by protecting against attacks that disrupt connectivity or signaling procedures. \\
\noindent$\bullet$~\textbf{Denial-of-Service Resistance:} Protects against resource exhaustion attacks.\\
\noindent$\bullet$~\textbf{Signaling Flood Protection:} Detects and mitigates signaling flood attacks.\\
\noindent$\bullet$~\textbf{Rogue Base Station Detection:} Identifies and blocks connections to fake eNodeBs/gNodeBs.\\
\noindent$\bullet$~\textbf{Physical Layer Attack Protection:} Defends against RF jamming, replay, and signal spoofing attacks.

\subsection*{E.7 Interworking Security}
Ensures secure and consistent behavior when UEs operate across different access technologies, network generations, and specification layers. \\
\noindent$\bullet$~\textbf{Specification Downgrade Protection:} Blocks fallback to insecure or legacy specifications.\\
\noindent$\bullet$~\textbf{Cross-Specification Interaction Security:} Prevents vulnerabilities caused by NAS, RRC, SBA, and IP interactions.\\
\noindent$\bullet$~\textbf{Specification Compatibility:} Ensures secure, consistent behavior across 4G, 5G, NSA, and SA deployments.

\subsection*{E.8 Threat Detection \& Logging}
Enables real-time anomaly detection and comprehensive monitoring to ensure system integrity.\\
\noindent$\bullet$~\textbf{Cross-Layer Threat Detection:} Monitors interactions across NAS, RRC, SBA, and user-plane layers.\\
\noindent$\bullet$~\textbf{Logging \& Auditing:} Maintains secure, tamper-proof logs for incident analysis and forensic investigations.\\
\noindent$\bullet$~\textbf{Timer Behavior Verification:} Validates specification timers to prevent potential exploitation.\\
\noindent$\bullet$~\textbf{Counter Behavior Verification:} Ensures NAS, RRC, and PDCP counters are strictly monotonic, unique, and correctly synchronized between UE and network. Detects and prevents counter wrap-around, reuse, or desynchronization.

\subsection*{E.9 Regulatory Compliance}
Ensures adherence to legal, regulatory, and industry security requirements while protecting against fraudulent activities.\\ 
\noindent$\bullet$~\textbf{Regulatory Compliance:} Enforces lawful intercept, data protection, and privacy regulations.\\
\noindent$\bullet$~\textbf{Billing \& Fraud Protection:} Prevents manipulation of charging data and unauthorized resource usage.

\section{Test Case Example}
\label{append:testcase}
Table~\ref{tab:replay-2073} presents the example test case, which is designed to validate the \texttt{service integrity} violation identified in Node 2073.

\end{document}